\documentclass[12pt,aps,prb,preprint]{revtex4}   
\usepackage{graphicx}   % for figures
\usepackage{amsfonts}
\usepackage{amsmath}    % need for subequations
\usepackage{amssymb}
\usepackage{bm}
\usepackage{bbm}
\usepackage{fancyhdr}
\usepackage{textcomp}
\usepackage{mathrsfs}
\usepackage{enumerate}
\usepackage{animate}

\begin{document}

%%%%%%%%%%%%%%%%%%%%%%%%%%%%%%%%%%%%%%%%%%%%%
\title{
Magic mass ratios of complete energy-momentum transfer 
\\
in one-dimensional elastic three-body collisions
}
%%%%%%%%%%%%%%%%%%%%%%%%%%%%%%%%%%%%%%%%%%%%%
\author{June-Haak Ee}
\affiliation{Department of Physics, Korea University, Seoul 136-713, Korea}
\author{Jungil Lee}

\email{jungil@korea.ac.kr}   %optional
\affiliation{Department of Physics, Korea University, Seoul 136-713, Korea}

%%%%%%%%%%%%%%%%%%%%%%%%%%%%%%%%%%%%%%%%%%%%%
\date{\today}
%%%%%%%%%%%%%%%%%%%%%%%%%%%%%%%%%%%%%%%%%%%%%

\begin{abstract}
We consider the one-dimensional scattering of two identical blocks of mass $M$ that exchange energy and momentum via elastic collisions
with an intermediary ball of mass $m=\alpha M$. Initially, one block is incident upon the ball
with the other block at rest.  For $\alpha<1$, 
the three objects will make multiple collisions with one another.
In our analysis, we construct a Euclidean vector $\textbf{V}_n$
whose components are proportional to the velocities of the objects.
Energy-momentum conservation then requires a covariant recurrence 
relation for $\textbf{V}_n$ that transforms like
a pure rotation in three dimensions. The analytic solutions of
the terminal velocities result in a remarkable prediction for values of $\alpha$, in cases where the initial energy 
and momentum of the incident block are completely transferred to 
the scattered block. We call these values for $\alpha$ ``magic mass ratios.''
\end{abstract}

\maketitle

%%%%%%%%%%%%%%%%%%%%%%%%%%%%%%%%%%%%%%%%%%%%%
\section{Introduction\label{sec:intro}}
%%%%%%%%%%%%%%%%%%%%%%%%%%%%%%%%%%%%%%%%%%%%%

If two identical particles make a head-on collision,
they exchange energy and momentum in any reference 
frame. \cite{footnote1} In a frame where the target is initially 
at rest, the target can acquire all of the energy and momentum of the incident object, causing the incident object to come to rest.  Such a complete energy-momentum transfer, however, is rarely observed in many-body 
collisions.
A specific example is Newton's cradle, which consists of
a series of identical pendula that are aligned
on a horizontal level.  Newton's cradle can be understood as a series of head-on collisions of
two identical objects where small gaps
between each pair of adjacent pendula are assumed.
\cite{Hart:1968, Kerwin:1972, Gavenda:1997}

Although this small-gap model for Newton's cradle explains the observations,
the model fails if some of the adjacent pendula are in contact with each other, in which case
the process cannot be decomposed into a series of two-body collisions.
The reason for this failure is 
that the constraints of energy-momentum conservation
are not enough to determine the final velocities.
Additional conditions, such as force laws between objects, are required to obtain
a unique solution of the problem.
Various studies have been carried out 
to find an appropriate force law that governs 
the actual motions of the pendula. 
For example, the contact Hertz force of the form 
$F=-kx^{3/2}$ is such a phenomenological model. \cite{Gavenda:1997, Chapman:1960, Flansburg:1979,
Herrmann:1981, Hermann:1982}
Further applications of the contact Hertz force to various collision problems can be found in
Refs.~\citenum{Leroy:1985, Patricio:2004, Hessel:2006, Cross:2007, Cross:2008}.

In a previous work, we studied the bouncing of a block against a rigid wall
through one-dimensional multiple elastic collisions with a ball sandwiched between them.
\cite{Ee:2012}
Based on the assumption that 
each collision is instantaneous, as in the small-gap model,
we obtained the unique analytic solutions to the complete trajectories 
of the block and the ball. 
By taking the continuum limit of the block trajectory,
we have shown that the effective force carried by the ball is proportional to $1/r^3$,
where $r$ is the distance between the block and the wall. This is consistent with previous results based on the differential
equation that can be derived by taking the continuum limit of energy-momentum conservation.
\cite{De:1997, Sinai:1999}

\begin{figure}[!h]
\begin{center}
\includegraphics[width=6.5cm]{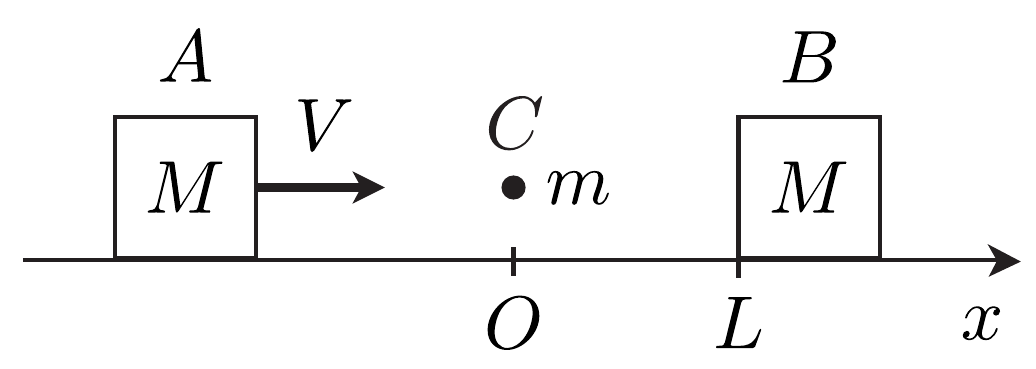}
\end{center}
\vspace{-1cm}
\caption{
The initial condition of the model system.
Block $A$ with initial speed $V$ is approaching
ball $C$ and block $B$ that are at rest at
$x=0$ and $L$, respectively.
Both blocks have mass $M$ and $m=\alpha M$ is the
mass of the ball, with $\alpha<1$.
\label{fig:system}} 
\end{figure}

\begin{figure}[!h]
\begin{center}
\includegraphics[width=10.5cm]{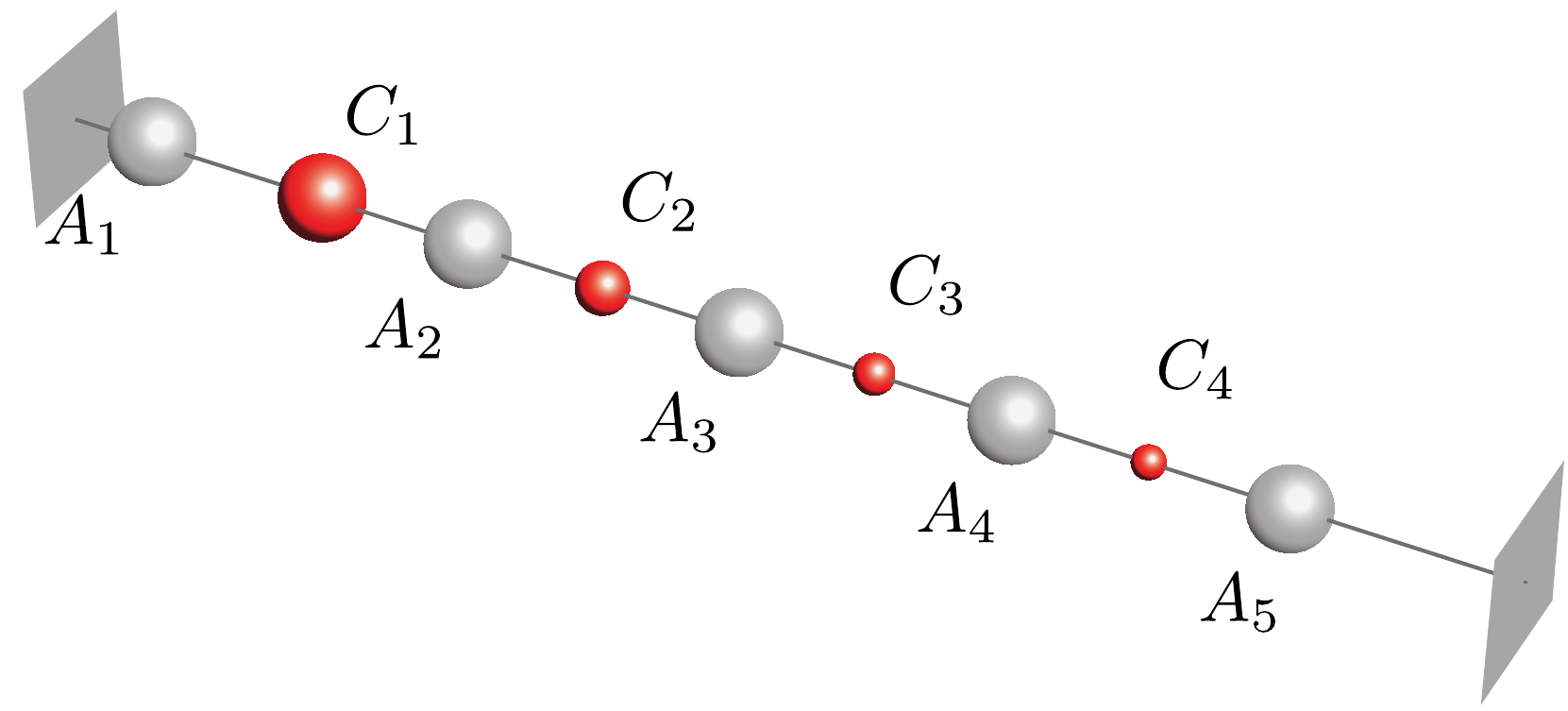}
\caption{\label{fig:system2}
A generalized version of Newton's cradle
consisting of beads on a horizontal frictionless straight wire.
There are five identical ``blocks'' $A_k$ of mass $M$
on a straight wire and a ``ball'' 
$C_k$ of mass $m_k=\alpha_k^{\textrm{magic}}M$
between each set of blocks.
Initially, all blocks are at rest except $A_1$, which is moving towards $C_1$.
After $A_k$ makes the $k$th collision with $C_k$, $A_k$ stops and
$C_k$ makes the $k$th collision with $A_{k+1}$. Then $C_k$ stops and
$A_{k+1}$ carries the complete amount of the incident momentum of $A_{k}$.
This process continues until $A_5$ carries the initial momentum of $A_1$.
After that, $A_5$ bounces back against the wall and the process repeats
in the reverse direction.  (Enhanced online)
[URL: http://dx.doi.org/10.1119/1.4897162.1]
} 
\end{center}
\end{figure}

In this paper, we generalize the model system of Ref.~\citenum{Ee:2012}
to a simple three-body system.  Here we consider the scattering of two identical blocks of mass $M$ through
multiple elastic collisions with a ball of mass $m=\alpha M$ sandwiched
between the blocks. As shown in Fig.~\ref{fig:system}, the ball $C$ and the target
block $B$ are initially placed at locations $x=0$ and $L$, respectively, and
the block $A$ is incident to the target with initial velocity $V$.
If $\alpha<1$, then the system experiences
multiple collisions that can be understood as a series of two-body 
collisions and, therefore, the velocities of all objects are uniquely
determined. To aid in visualizing the discussion below, the reader is encouraged
to view the animation of a series of such collisions in the online video linked to Fig.~\ref{fig:system2}.

In our analysis, we construct a Euclidean vector whose components are proportional to 
the velocities of the objects.
With such, the energy-momentum conservation constraints reduce to a pure rotational
transformation under which the magnitude of the Euclidean vector is invariant.
This covariant approach is equivalent to the billiard-ball mapping 
approach in 
Refs.~\citenum{Sinai, Kozlov, Gutkin:1996, Tabachnikov, Redner:2004}, that reformulates the collision problem into a problem
of projective geometry.
While we were interested in the trajectories of the objects and the effective
force in Ref.~\citenum{Ee:2012}, here
we focus on the energy-momentum transfer from the incident block $A$
to the scattered block $B$. 
As a result, we find all possible values for the
\textit{magic mass ratios} $\alpha=\alpha^{\textrm{magic}}$ 
at which the energy and momentum of the incident block are completely transferred
to the scattered block.

The chain collisions of multiple pendula
in series have been studied previously by
Hart {\it et al.} \cite{Hart:1968} and by Kerwin. \cite{Kerwin:1972}
However, the pendula studied in these references are arranged
in mass order so that the complete energy-momentum transfer from the incident pendulum
to the target pendulum at the other end is achieved only if all of the pendula are of equal mass, as in Newton's cradle.
Redner considered a similar system consisting of two identical cannonballs approaching 
an initially stationary ping-pong ball.~\cite{Redner:2004}
In that study, the author focuses mainly on deriving a simple relation 
between the elastic collision and a corresponding billiard system,
which helps to determine the total number of collisions of the system.
Thus, to our best knowledge, the derivation of all possible magic mass ratios $\alpha^{\textrm{magic}}$ 
for which complete energy-momentum transfer is realized is new.

This paper is organized as follows. 
In Sec.~\ref{sec:model}, we describe the model system and provide 
the definitions of kinematic variables that we use throughout
the paper.
In Sec.~\ref{sec:v}, we construct the covariant
recurrence relation for 
the velocity sequences of the objects and compare this approach with
the billiard-ball mapping. Our results for the terminal velocities and 
kinematic variables including the magic mass ratios
are given in Sec.~\ref{sec:result}. 
Finally, we offer our conclusions in Sec.~\ref{sec:concl} and provide some technical
details in the Appendices.

%%%%%%%%%%%%%%%%%%%%%%%%%%%%%%%%%%%%%%%%%%%%%%
\section{Model, terminology, and conjectures\label{sec:model}}
%%%%%%%%%%%%%%%%%%%%%%%%%%%%%%%%%%%%%%%%%%%%%%

In this section, we describe the model system and define kinematic
variables that we use throughout the paper. 
We also discuss some conjectures regarding 
the relationship among special values 
for mass ratios.

%--------------------------------------------------
\subsection{Model\label{subsec:model}}
%--------------------------------------------------Sect.

As shown in Fig.~\ref{fig:system}, the model system consists of 
two identical blocks $A$ and $B$, each of mass $M$, and a ball $C$ 
with mass $m=\alpha M$, all aligned on the $x$-axis.
Initially, $C$ and $B$ are at rest
at $x=0$ and $L$, respectively, and block $A$ is at $x<0$
and moving toward ball $C$ with speed $V$.  At time $t=0$,
block $A$ collides with the ball.
We assume that all of the collisions are (completely)
elastic and ignore friction.  If the mass of $C$ is
equal to or larger than the mass of
the blocks ($\alpha\ge 1$), then $C$ will collide with
each block only once.  But if $C$ is less massive than
the blocks ($\alpha<1$), then it moves back and
forth to make multiple collisions with the blocks.
In either case, because each collision is governed by
a repulsive force, $A$ must decelerate and $B$
must accelerate.  In this paper, we are interested
in the situation where $A$ and $B$ interact
through multiple elastic collisions with $C$ (i.e. $\alpha<1$).

%--------------------------------------------------
\subsection{Terminology\label{subsec:term}}
%--------------------------------------------------

We use an integer
$n$ to identify the sequence of collisions of the ball with a block.
We denote by $P_n$ the $n$th collision point between objects $A$ 
and $C$, and by $Q_n$ the $n$th collision point between objects $B$ and $C$.
The velocities of the three objects are defined
as follows:  $a_n$ and $c_n$ are
the velocities of $A$ and $C$, respectively,
immediately after the collision at $P_n$. Likewise,
$b_n$ and $c^{\,\prime}_n$ are
the velocities of $B$ and $C$, respectively,
immediately after the collision at $Q_n$.
During the ball's motion from $P_n$ to $Q_n$, the velocity of $C$ 
is fixed at $c_n$; between $Q_n$ and $P_{n+1}$, the velocity of 
$C$ is $c'_n$.

Because $\alpha$ is a finite number, the total number 
of collisions between $C$ and a block is also finite. We call
$N_A$ and $N_B$ the total number of collisions between
blocks $A$ and $B$ and the ball, respectively.
After these final collisions, blocks
$A$ and $B$
reach their terminal velocities
$a_t=a_{N_A}$ and
$b_t=b_{N_B}$, respectively. Conservation
of (linear) momentum
determines the terminal velocity of $C$ as
$c_t=(V-a_t-b_t)/\alpha$.

If $\alpha$ is decreased, then $N_A$ and $N_B$ will
increase by unity at certain values of $\alpha$. We call $\alpha_k$ and
$\beta_k$ the $k$th \textit{threshold mass ratios} of $A$ and $B$, which
are the minimum values for $\alpha$ with
$N_A=k$ and $N_B=k$ for $k\ge 1$, respectively.
Therefore, the values for $N_A$ and $N_B$ are determined
by the mass ratio $\alpha$ as
%-------------
%\begin{subequations}
\begin{equation}
\label{NA-range}
N_A=k\quad\textrm{for}\quad \alpha_{k}\le\alpha< \alpha_{k-1}
\end{equation}
and
\begin{equation}
\label{NB-range}
N_B=k\quad\textrm{for}\quad \beta_{k}\le\alpha< \beta_{k-1},
\end{equation}
%\end{subequations}
%-------------
where we set $\alpha_0=\infty$ and $\beta_0=\infty$.

In a many-body, fixed-target collision, it is not easy to determine the masses
so that the kinetic energy of the incident particle is
completely transferred to a single scattered particle
(leaving the incident particle at rest). 
However, in some situations,
such as with
Newton's cradle or the situation we consider here,
it is possible to solve for the masses.
There are two classes of critical mass ratios,
the \textit{magic mass ratio} and the \textit{deficient mass ratio}.
We denote by $\alpha_k^{\textrm{magic}}$ 
the $k$th magic mass ratio at which both $A$ and $C$ stop moving
and $B$ carries the complete energy and momentum of the incident block.
Between any two consecutive magic mass ratios 
$\alpha_k^{\textrm{magic}}$ and 
$\alpha_{k+1}^{\textrm{magic}}$, there will be a mass ratio for which the
energy-momentum transfer is the least efficient. 
We call this value the $k$th deficient mass ratio 
$\alpha_k^{\textrm{deficient}}$.

%--------------------------------------------------
\subsection{Conjectures $\bm{\alpha_k^\textrm{magic}=\alpha_k}$ 
and $\bm{\alpha_k^\textrm{deficient}=\beta_k}$\label{subsec:conjecture}}
%--------------------------------------------------
Before working through the details, we can make an educated guess that 
$\alpha_k^\textrm{magic}=\alpha_k$.
Let us suppose that there exists $\alpha_k^{\textrm{magic}}$
where the scattered block takes all of the kinetic energy
of the incident block. Thus, the momentum transfer from
block $A$ to block $B$ is at a maximum.
In this case, both $A$ and $C$ must
remain at rest after the final collision.
Such a configuration is feasible only
if $A$ stops at $P_k$ ($a_k=0$),
$C$ stops at $Q_k$ ($c_k'=0$), and $B$ carries
the initial velocity of the incident block $A$ ($b_k=V$).
In fact, these conditions require\cite{footnote1x}
$c_k=2V/(1+\alpha_k^{\textrm{magic}})$, and 
also lead to $N_A=N_B=k$.
Next, we take 
$\alpha=\alpha_k^{\textrm{magic}}-\varepsilon$, which is infinitesimally smaller
than $\alpha_k^{\textrm{magic}}$ $(\varepsilon\to 0^+)$.
Then the momentum transfer between $A$ and $B$ that
is mediated by $C$ must be infinitesimally
less than when $\alpha = \alpha_k^{\textrm{magic}}$, and the
ball must carry a surplus of energy after the last collision, and this will
lead to an additional collision with block $A$.

Similarly, we might guess that at $\alpha = \beta_k$ the final kinetic energy of the ball reaches a
local maximum. We shall find that these conjectures are in fact true:
$\alpha_k^{\rm magic} = \alpha_k$ and $\alpha_k^{\rm deficient} = \beta_k$ for all $k$.

%%%%%%%%%%%%%%%%%%%%%%%%%%%%%%%%%%%%%%%%%%%%%
\section{Analytic Computation in Covariant Approach\label{sec:v}}
%%%%%%%%%%%%%%%%%%%%%%%%%%%%%%%%%%%%%%%%%%%%%

In this section, we use energy and momentum conservation to
develop a recurrence relation for a column vector $\textbf{V}_n$
constructed from the velocities of blocks $A$ and $B$ and ball $C$.
This approach is compared with an alternative method---the billiard-ball
mapping of Redner---that can be applied to
a similar problem. \cite{Redner:2004}

%--------------------------------------------------
\subsection{Construction of a Euclidean vector
in terms of velocities\label{subsec:Euclid}}
%--------------------------------------------------

It is convenient to define the column-vector sequence
$\textbf{V}_n$ as
%-------------
\begin{equation}
\label{def:Vn}
\textbf{V}_n\equiv
\begin{pmatrix}
a_n\\b_n\\\sqrt{\alpha}c^{\,\prime}_n
\end{pmatrix},
\end{equation}
%-------------
where the scaling factor $\sqrt{\alpha}$ of the third component
is introduced to simplify the following analyses. 
The magnitude of the three-dimensional Euclidean vector $\textbf{V}_n$ is defined by
$|\textbf{V}_n|\equiv \sqrt{\textbf{V}_n^2}$, where $\textbf{V}_n^2\equiv(V_n)_1^2+(V_n)_2^2+(V_n)_3^2$
and $(V_n)_i$ is the $i$th Cartesian component of $\textbf{V}_n$; namely,
$(V_n)_1=a_n$,
$(V_n)_2=b_n$, and
$(V_n)_3=\sqrt{\alpha}c_n^\prime$.
We shall see in the following section that energy conservation forces 
$|\textbf{V}_n|$ to be independent of $n$ and the transformation of 
$\textbf{V}_{n-1}$ into $\textbf{V}_{n}$ is linear.
This covariant transformation can be formulated by making
use of an orthogonal matrix $\mathcal{O}$.
(Some useful properties of orthogonal matrices are summarized
in Appendix~\ref{app:orthogonal}).

%===============================================================
\subsection{Covariant recurrence relation for $\textbf{V}_n$\label{sec:rec}}
%===============================================================

The collision at $P_n$ transforms the velocities
$a_{n-1}$ and $c^{\,\prime}_{n-1}$ into $a_n$ and $c_n$, respectively.
The collision at $Q_n$ transforms the velocities
$b_{n-1}$ and $c_{n}$ into $b_n$ and $c^{\,\prime}_n$, respectively.
Conservation of momentum and kinetic energy
in the collisions at $P_n$ and $Q_n$ requires
%-----------------------
%\begin{subequations}
\begin{eqnarray}
\label{eq:conservation}
%\begin{eqnarray}
%-----------------------
\label{eq:ac}
a_{n-1}+\alpha c^{\,\prime}_{n-1}&=&a_n+\alpha c_n,
\\
\label{eq:Euc-Pn}
a^2_{n-1}+\left(\sqrt{\alpha} c^{\,\prime}_{n-1}\right)^2&=&a_n^2+\left(\sqrt{\alpha} c_n\right)^2,
\\
\label{eq:bcp}
b_{n-1}+\alpha c_{n}&=&b_n+\alpha c^{\,\prime}_n,
\end{eqnarray}
and
\begin{equation}
\label{eq:Euc-Qn}
b^2_{n-1}+\left(\sqrt{\alpha} c_{n}\right)^2=
b_n^2+\left(\sqrt{\alpha} c^{\,\prime}_{n}\right)^2.
\end{equation}
%-----------------------
%\end{subequations}
%-----------------------
The sum of Eqs.~(\ref{eq:Euc-Pn}) and (\ref{eq:Euc-Qn})
yields the energy-conservation constraint under the $n$th 
cycle of
collisions at $P_n$ and $Q_n$:
\begin{equation}
\label{eq:energy-conservation-PnQn}
a^2_{n-1}+b^2_{n-1}
+\left(\sqrt{\alpha} c^{\,\prime}_{n-1}\right)^2
=
a^2_{n}+b^2_{n}
+\left(\sqrt{\alpha} c^{\,\prime}_{n}\right)^2.
\end{equation}
By coupling to the constraint (\ref{eq:ac}),
we can reduce Eq.~(\ref{eq:Euc-Pn}) for
the collision at $P_n$ into a linear form. In a similar manner,
Eq.~(\ref{eq:Euc-Qn}) for the collision at $Q_n$ also 
reduces into a linear form. The result is
%-----------------------
%\begin{subequations}
\begin{eqnarray}
%\label{eq:conservation-linear}
%\begin{eqnarray}
%-----------------------
\label{eq:Euc-Pn-linear}
a_{n-1}- c^{\prime}_{n-1}
&=&-a_n+c_n,
\\
\label{eq:Euc-Qn-linear}
b_{n-1}-c_{n\phantom{-1}}&=&
-b_n+c^{\,\prime}_{n}.
\end{eqnarray}
%-----------------------
%\end{subequations}
%-----------------------
The reduction in Eqs.~(\ref{eq:Euc-Pn-linear})
and \eqref{eq:Euc-Qn-linear} is 
allowed as long as 
$a_{n-1}\neq a_{n}$,
$b_{n-1}\neq b_{n}$,
$c^\prime_{n-1}\neq c_{n}$, and
$c_{n}\neq c^\prime_{n}$.
These requirements are always satisfied
because every collision changes the
velocity of each participant.

If we now employ the notation for the vector $\textbf{V}_n$
defined in Eq.~(\ref{def:Vn}), we find that
Eq.~(\ref{eq:energy-conservation-PnQn})
can be expressed as
%\begin{subequations}
%\label{eq:transform}
\begin{equation}
\label{eq:V2}
\textbf{V}_n^2=\textbf{V}_{n-1}^2.
\end{equation}
Similarly, after eliminating $c_n$ from Eqs.~(\ref{eq:ac}), (\ref{eq:bcp}),
(\ref{eq:Euc-Pn-linear}), and
(\ref{eq:Euc-Qn-linear}), we end up with three equations 
that can be written as
\begin{equation}
\label{eq:V-n}
\textbf{V}_n=\Lambda \textbf{V}_{n-1},
\end{equation}
%\end{subequations}
%------------------ 
where $\Lambda$ is a $3\times 3$ matrix 
that depends only on $\alpha$.
By applying the recurrence relations~(\ref{eq:V2})
and \eqref{eq:V-n}
repeatedly, we find that the length of vector $\textbf{V}_n$ is invariant
($\textbf{V}_n^2=\textbf{V}_0^2$) and
%------------------
\begin{equation}
\label{eq:V-n-0}
\textbf{V}_n=\Lambda^n \textbf{V}_0,
\end{equation}
%------------------
where $\textbf{V}_0=(V\, 0\,\, 0)^T$.

The constraints in Eqs.~(\ref{eq:V2})
and \eqref{eq:V-n} are equivalent to the covariant transformation
rule for a Euclidean vector $\textbf{V}\to \textbf{V}'=\mathcal{O} \textbf{V}$ that preserves 
its magnitude $|\textbf{V}'|=|\textbf{V}|$.
In a Cartesian coordinate system, the magnitude is defined by
$|\textbf{V}|=\left(\textbf{V}\cdot \textbf{V}\right)^{1/2}=\left(\sum_{i}V_{i}^2\right)^{1/2}$, where $V_i$ is the
$i$th Cartesian coordinate of $\textbf{V}$. Such a transformation also preserves any scalar
product of such vectors $\textbf{V}\cdot\textbf{W}=\sum_{i,j}V_i\delta_{ij}W_j=\sum_{i}V_iW_i$,
where the Kronecker delta $\delta_{ij}$ is the \textit{metric tensor} of
the Euclidean space. The invariance of the scalar product under this transformation
requires that the matrix $\mathcal{O}$ must be orthogonal;
stated another way, the inverse of
$\mathcal{O}$ is the same as its transpose: $\mathcal{O}^{-1}=\mathcal{O}^T$.
In addition, for any orthogonal matrix $\mathcal{O}$, $\det[\mathcal{O}]=\pm 1$.
An orthogonal matrix with $\det[\mathcal{O}]=+ 1$ can be expressed as 
a pure rotation, while an orthogonal matrix with $\det[\mathcal{O}]=-1$ can be
expressed as either a reflection or the product of a pure rotation and a reflection.
We call the formulation in Eqs.~(\ref{eq:V2})
and \eqref{eq:V-n}
the \textit{covariant} approach. 

By generalizing the method in Ref.~\citenum{Ee:2012} for two-body
multiple collisions against a rigid wall, we develop a systematic
way to determine $\Lambda$ and $\Lambda^n$.
As is stated in Appendix~\ref{app:orthogonal}, the orthogonal
matrix $\Lambda$ represents a pure rotation  about 
an axis $\hat{n}$ by a finite angle $\psi$
in three dimensions.  Thus, according to Eq.~(\ref{eq:V-n-0}),
$\textbf{V}_n$ can be obtained by rotating $\textbf{V}_0$
about the same axis $\hat{n}$ by $n\psi$.
Because $\alpha$ is the only variable that
involves the transformation of $\textbf{V}_n$, 
both $\hat{n}$ and $\psi$ depend
only on $\alpha$.

%===============================================================
\subsection{Comparison with billiard-ball mapping\label{sec:comparison}}
%===============================================================

In principle, the covariant approach summarized in Eqs.~(\ref{eq:V2}) 
and (\ref{eq:V-n})
is equivalent to the billiard-ball mapping described
by Redner in Ref.~\citenum{Redner:2004}. Redner also rescaled the velocity
by multiplying the square root of its mass 
as we did in constructing the Euclidean vector $\textbf{V}_n$.
The crucial constraint in Redner's billiard-ball mapping is that the scalar product
$(\sqrt{m_1},\sqrt{m_2})\cdot (w_1,w_2)$ is conserved. Here, 
$m_i$ is the mass of the $i$th particle in a two-body collision
and $w_i=\sqrt{m_i}v_i$ is the rescaled velocity $v_i$ of particle $i$:
$(\sqrt{m_1},\sqrt{m_2})\cdot (w_1,w_2)=%
(\sqrt{m_1},\sqrt{m_2})\cdot (w'_1,w'_2)$,
where the primed variable is used
for a particle after a collision.
This constraint originates from the momentum conservation law
in Eqs.~(\ref{eq:ac}) and (\ref{eq:bcp}) and is equivalent to the covariant 
transformation rule of Eq.~(\ref{eq:V-n}). Geometrically, the conservation of
the projection of the vector $(w_1,w_2)$ onto
the vector $(\sqrt{m_1},\sqrt{m_2})$ implies that the transformation
of $(w_1,w_2)$ into $(w'_1,w'_2)$ must be related to a rotation about
an axis parallel to $(\sqrt{m_1},\sqrt{m_2})$ and/or a reflection of
the component perpendicular to $(\sqrt{m_1},\sqrt{m_2})$.
Therefore, the covariant approach
must be equivalent to the billiard-ball
mapping approach. Redner then reformulated the problem into a problem of projective
geometry.

The merit of the covariant approach is that the corresponding
computation techniques are available in standard textbooks 
on classical mechanics or mathematical physics, such as 
Refs.~\citenum{Marion} and \citenum{Arfken}.
An extension of the covariant approach into a relativistic collision
is even feasible at the level of undergraduate physics majors.
A generalization of the billiard-ball mapping of a
projective geometry
into that in the Minkowski space
needs additional work.

%===============================================================
\subsection{Determination of $\bm{\Lambda}$\label{sec:Lambda}}
%===============================================================

The recurrence relations~(\ref{eq:V2})
and \eqref{eq:V-n}
represents the transformation of the velocities after two collisions at
$P_n$ and $Q_n$. Because each collision is elastic, there exist 
orthogonal (\textit{collision}) matrices---$\Gamma_A$ for the collision at $P_n$ and 
$\Gamma_B$ for that at $Q_n$---that preserve the magnitude of $\textbf{V}_n$.
In this case, $\Lambda$ must be expressed as
%------------------
\begin{equation}
\label{eq:Lam-G-G}
\Lambda=\Gamma_B\Gamma_A.
\end{equation}
%------------------

Here, $B$ does not participate in the collision at $P_n$ and $A$
is independent of the collision at $Q_n$ so that $\Gamma_A$ and
$\Gamma_B$ maintain the velocities of $B$ and $A$, respectively. Thus, $\Gamma_A$
and $\Gamma_B$ are intrinsically two-dimensional linear transformations.
The matrix $\Gamma_i$ for the elastic collision of 
one-dimensional two-body scattering has the determinant $-1$
(see, for example, Ref.~\citenum{Ee:2012}).
Therefore, each collision matrix can be
parametrized by the product of a pure rotation and a reflection so that
$\textrm{det}[\Gamma_A]=\textrm{det}[\Gamma_B]=-1$.
By making use of the properties of $3\times 3$ orthogonal matrices 
summarized in Appendix~\ref{app:orthogonal},
we find the useful parametrizations
%------------------
%\begin{subequations}
\begin{equation}
\Gamma_A=\lambda_2(-\theta)\mathbbm{P}_3
\end{equation}
and
\begin{equation}
\Gamma_B=\lambda_1(\theta)\mathbbm{P}_3,
%------------------.
\end{equation}
%\end{subequations}
%------------------
where 
$\lambda_i(\theta)$ are rotation matrices about the $i$th Cartesian axis,
$\mathbbm{P}_3=\textrm{diag}[1,1,-1]$ is a reflection
matrix, and $\theta$ is given by
%------------------
\begin{equation}
\label{theta-txt}
\theta=2 \arctan \sqrt{\alpha}.
\end{equation}
%------------------
We then find the explicit form of the matrix $\Lambda$ as
%---------------------
\begin{equation}
\label{eq:Lambda}
%---------------------
\Lambda=
\begin{pmatrix}
\displaystyle{\frac{1-\alpha}{1+\alpha}}&
0&
\displaystyle{\frac{2\sqrt{\alpha}}{1+\alpha}}
\\[10 pt]
\displaystyle{\frac{4\alpha}{(1+\alpha)^2}}&
\displaystyle{\frac{1-\alpha}{1+\alpha}}&
\displaystyle{-\frac{2(1-\alpha)\sqrt{\alpha}}{(1+\alpha)^2}}
\\[12 pt]
\displaystyle{-\frac{2(1-\alpha)\sqrt{\alpha}}{(1+\alpha)^2}}&
\displaystyle{\frac{2\sqrt{\alpha}}{1+\alpha}}&
\displaystyle{\frac{(1-\alpha)^2}{(1+\alpha)^2}}
\end{pmatrix}.
%---------------------
\end{equation}
%---------------------
An explicit computation shows that
$\Lambda^T=\Lambda^{-1}$ (i.e. $\Lambda$ is orthogonal).
A detailed procedure to compute $\Lambda$ is given in
Appendix~\ref{app:Vn}. 

%===============================================================
\subsection{Computation of $\bm{\Lambda^n}$\label{sec:Lambda-n}}
%===============================================================

Direct computation of $\Lambda^n$ by repeated multiplications of $\Lambda$
is not trivial. However, the analysis can be greatly simplified if we make use
of the identity in Eq.~(\ref{eq:Lam-G-G}) and
the parametrization
$\lambda_{\hat{n}}(\phi)=
R\,\lambda_3(\phi)R^T$,
where $R$ is an orthogonal matrix
(see Appendix~\ref{app:orthogonal}).

Given that $\textrm{det}[\Gamma_A]=\textrm{det}[\Gamma_B]=-1$,
we know that 
$\textrm{det}[\Lambda]=\textrm{det}[\Gamma_A]\textrm{det}[\Gamma_B]=+1$. 
Therefore,
$\Lambda$ can be parametrized by a pure rotation $\Lambda=R\lambda_3(\psi)R^{T}$,
where the angle of rotation $\psi$ must be a function of $\alpha$ only.
Because $R$ is orthogonal, it is straightforward to find that
%---------------------
\begin{equation}
\label{eq:Lambda-n-final-txt} 
%---------------------
\Lambda^n=R\lambda_3(n\psi)R^T,
%---------------------
\end{equation}
%---------------------
where we have used the fact that $[\lambda_3(\psi)]^n=\lambda_3(n\psi)$.
According to the expression for $\Lambda^n$ in 
Eq.~(\ref{eq:Lambda-n-final-txt}), it is manifest that an additional
cycle of collisions at $P_n$ and $Q_n$ merely increases the phase
by a constant $\psi$ about the same axis of rotation.

A detailed derivation of $\Lambda^n$ is provided in Appendix~\ref{app:lam-n};
here we simply quote the
$\alpha$ dependence of $\psi$ and $R$:
%-------------
%\begin{subequations}
\begin{equation}
\label{def:psi}
\psi= 2\arctan\sqrt{\alpha(2+\alpha)}
\end{equation}
and
\begin{equation}
R=
\begin{pmatrix}
\displaystyle{-\frac{1}{\sqrt{2}}}&
\displaystyle{-\sqrt{\frac{\alpha}{2(2+\alpha)}}}&
\displaystyle{\frac{1}{\sqrt{2+\alpha}}}
\\[10 pt]
\displaystyle{\frac{1}{\sqrt{2}}}&
\displaystyle{-\sqrt{\frac{\alpha}{2(2+\alpha)}}}&
\displaystyle{\frac{1}{\sqrt{2+\alpha}}}
\\[10 pt]
0&
\displaystyle{\sqrt{\frac{2}{2+\alpha}}}&
\displaystyle{\frac{\sqrt{\alpha}}{\sqrt{2+\alpha}}}
\end{pmatrix}.
\end{equation}
%\end{subequations}
%-------------
As with $\Lambda$, an explicit computation confirms that $R$ is orthogonal
($R^T = R^{-1}$).  Having found $R$, we know that the
third column of $R$
is the unit vector $\hat{n}$ that is
parallel to the axis of rotation for $\Lambda$. We find that $\hat{n}$ is parallel
to $(\sqrt{M_A}\;\sqrt{M_B}\;\sqrt{M_C})^T$, where $M_i$ is the mass of
$i=A$, $B$, and $C$, supporting our earlier argument that the covariant approach
is equivalent to the billiard-ball mapping approach.

%%%%%%%%%%%%%%%%%%%%%%%%%%%%%%%%%%%%%%%%%%%%%
\section{Results\label{sec:result}}
%%%%%%%%%%%%%%%%%%%%%%%%%%%%%%%%%%%%%%%%%%%%%

In this section, we find the general terms of the
velocity sequences $a_n$, $b_n$, $c_n$, and $c^{\,\prime}_n$ 
by solving the recurrence relation for $\textbf{V}_n$.
By making use of these analytic solutions, 
we calculate $N_A$ and $N_B$, and obtain the threshold mass ratios
$\alpha_k$ and $\beta_k$.
We then determine the terminal velocities of
the objects and show that
the magic mass ratio $\alpha_k^\textrm{magic}$ and
the deficient mass ratio
$\alpha_k^\textrm{deficient}$ are identical to
the threshold mass ratios $\alpha_k$ and $\beta_k$, respectively.
Lastly, we discuss possible modifications of
our predictions for inelastic collisions.

Substituting the initial condition $\textbf{V}_0=(V\, 0\,\, 0)^T$ and
$\Lambda^n$ in Eq.~(\ref{eq:Lambda-n-final-txt}) into Eq.~(\ref{eq:V-n-0}), we can determine $\textbf{V}_n$.
The value of $c_n$ can be computed by substituting $a_{n-1}$, 
$a_{n}$, and $c^{\,\prime}_{n-1}$ into Eq.~(\ref{eq:ac}). 
The general terms for the velocity sequences are obtained as
%-------------
%\begin{subequations}
%\label{eq:abccp-final}
\begin{eqnarray}
\label{eq:a-final}
a_n&=&V_{\rm{CM}}
\left\{1+\frac{\cos (n\psi)}{\cos(\psi/2)}\right\},
\\
\label{eq:b-final}
b_n&=&V_{\rm{CM}}
\left\{1-\frac{\cos[(n+\tfrac{1}{2})\psi]}{\cos(\psi/2)}\right\},
\\
\label{eq:c-final}
c_n&=&V_{\rm{CM}}
\left\{1+\frac{\sin[(n-\tfrac{1}{4})\psi]}{\sin(\psi/4)}\right\},
\end{eqnarray}
and
\begin{equation}
\label{eq:c'-final}
c^{\,\prime}_n=V_{\rm{CM}}
\left\{1-\frac{\sin[(n+\tfrac{1}{4})\psi]}{\sin(\psi/4)}\right\},
\end{equation}
%\end{subequations}
%-------------
where $V_{\rm{CM}}$ is the velocity of the center-of-mass, given by
%-------------
\begin{equation}
V_{\rm{CM}}=\frac{V}{2+\alpha}=\frac{V\cos(\psi/2)}{1+\cos(\psi/2)}.
\end{equation}
%-------------
Note that $(m+2M)V_{\rm{CM}}=MV$ is the total linear momentum of the system.
If we make a Galilean transformation into the center-of-momentum frame, where
the center-of-mass is at rest, then the unity term
in each pair of braces 
in Eqs.~(\ref{eq:a-final})--\eqref{eq:c'-final} vanishes.
In deriving these results,
we have used the values for trigonometric functions at
$\psi$, $\psi/2$, and $\psi/4$ that are
listed in Table~\ref{table:trig}.
We also need to evaluate the trigonometric
functions at
$n\theta$, $(n+\tfrac{1}{2})\theta$, $n\psi$, $(n+\tfrac{1}{2})\psi$,
and $(n+\tfrac{1}{4})\psi$.
In Appendix~\ref{sec:tri}, we summarize a way to evaluate
these functions in terms of $\alpha$.

Let us summarize properties of the solutions listed in 
Eqs.~(\ref{eq:a-final})--\eqref{eq:c'-final}. For $n=0$, the velocities satisfy the initial conditions
$a_0=V$, $b_0=c_0^\prime=0$. As $n$ increases, $a_n$ decreases
and $b_n$ increases. The reason is that the impulse on $A$ in each collision
is along the $-x$ axis and the impulse on $B$ in each collision
is along the $+x$ axis.
The terminal velocities can be obtained once we know
the total numbers of collisions $N_A$ and $N_B$, which are
calculated in Appendix~\ref{sec:N-app}; the results are:

%-------------
\begin{equation}
\label{eq:NANB-ans-txt}
N_A
=
\begin{cases}
N_B=\lceil\pi/\psi\rceil-1&
{\rm for}~~0<\mathfrak{s}(\pi/\psi)\le 1/2,
\\
N_B+1=\lceil\pi/\psi\rceil,&
\textrm{otherwise}.
\end{cases}
\end{equation}
%-------------
Here, the ceiling function $\lceil x\rceil$ and
the  sawtooth function $\mathfrak{s}(x)$ are defined in Appendix 
\ref{sec:N-app}.
Using this result, along with Eqs.~(\ref{NA-range}) and \eqref{NB-range},
we find that
%\begin{subequations}
%\label{eq:NaNbatcritical}
\begin{eqnarray}
\label{eq:Naatcritical}
N_A&=&N_B=k=\frac{\pi}{\psi}-\frac{1}{2}\quad\textrm{if}\quad\alpha=\alpha_k,
\\
\label{eq:Nbatcritical}
N_A-1&=&N_B=k=\frac{\pi}{\psi}-\,1 \quad\textrm{if}\quad\alpha=\beta_k.
\end{eqnarray}
%\end{subequations}
At $\alpha=1$,
$\psi=2\pi/3$ and $N_A=N_B=1$.
As $\alpha$ decreases, $\psi$ also decreases.
At certain mass ratios, $\alpha=\alpha_k-0^+$ and $\alpha=\beta_k-0^+$,
$N_A$ and $N_B$ increase by unity, respectively.
The equalities in Eqs.~(\ref{NA-range}) and \eqref{NB-range} are determined by the step functions
in Eq.~(\ref{eq:NANB-ans-txt}). 
Note that the threshold mass ratios are ordered as
$\cdots<\beta_{2}<\alpha_2<\beta_1<\alpha_1=1$.

By making use of the analytic solutions given in 
Eqs.~(\ref{eq:a-final})--\eqref{eq:c'-final}, we can find every pair of $k$ and $\psi$
that guarantee the terminal velocities
satisfy the conditions $\alpha=\alpha_k^\textrm{magic}$ and $\alpha=\alpha_k^\textrm{deficient}$,
respectively;
this verifies the existence of both $\alpha_k^\textrm{magic}$ and $\alpha_k^\textrm{deficient}$.
The corresponding proofs for the existence of $\alpha_k^\textrm{magic}$ and $\alpha_k^\textrm{deficient}$
are given in Appendices \ref{app:magic} and \ref{app:deficient}, respectively.

The resultant values for the critical mass ratios can be 
obtained from Eqs.~(\ref{def:psi}) and (\ref{eq:Naatcritical})--\eqref{eq:Nbatcritical} as
%-------------
%\begin{subequations}
%\label{exact:akbk}
\begin{equation}
\label{exact:ak}
\alpha_k^{\textrm{magic}}=\alpha_k=-1+\sec\left(\frac{\pi}{2k+1}\right)
\end{equation}
and
\begin{equation}
\label{exact:bk}
\alpha_k^{\textrm{deficient}}=\beta_k=-1+\sec\left[\frac{\pi}{2(k+1)}\right],
\end{equation}
%\end{subequations}
%-------------
for $k\ge1$. At $\alpha=\alpha_k^{\textrm{magic}}$,
$\psi=\pi/(k+\tfrac{1}{2})$ and at $\alpha=\alpha_k^{\textrm{deficient}}$,
$\psi=\pi/(k+1)$.
The first few values are
$\alpha_{1}=1$,
$\beta_1=\sqrt{2}-1\approx 0.414$,
$\alpha_{2}=\sqrt{5}-2\approx 0.236$, and
$\beta_2=(2-\sqrt{3})/\sqrt{3}\approx 0.155$.

\begin{figure}[!h]
\begin{center}
\includegraphics[width=7.5cm]{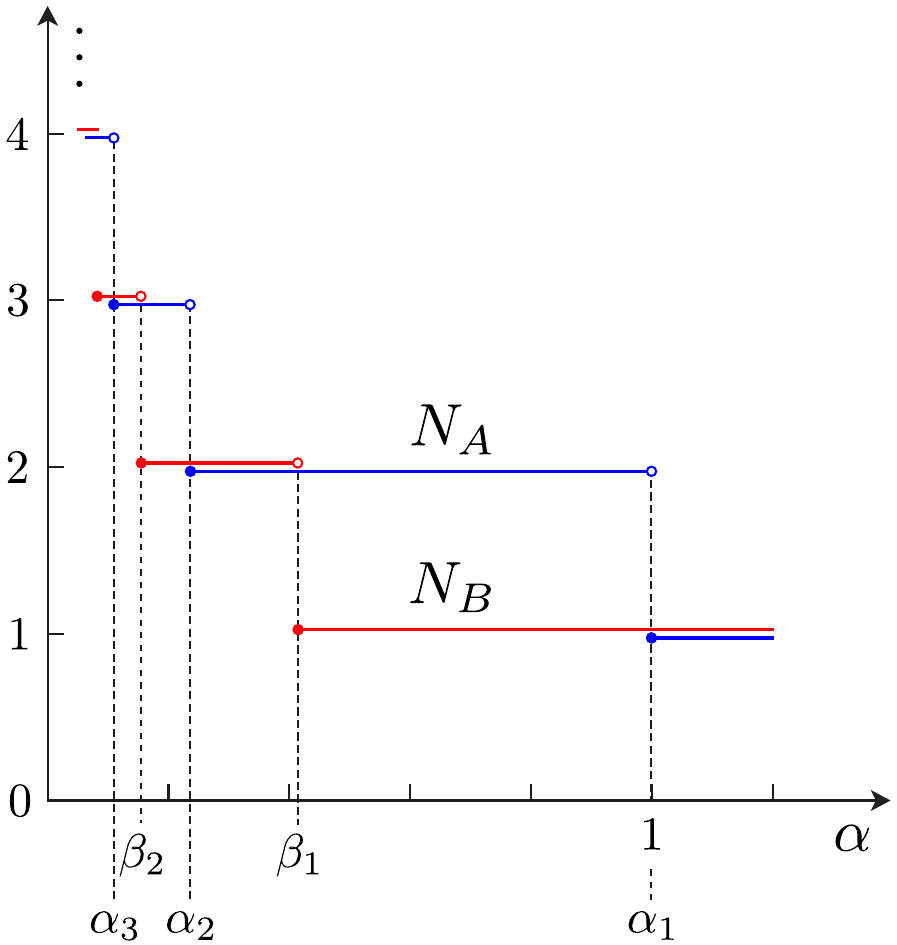}\\
\caption{\label{fig:NaNb}
The number of collisions $N_A$ and $N_B$ as functions of $\alpha$.
The threshold mass ratios
$\alpha_k$ and $\beta_{k}$ are the minimum values of $\alpha$
to have $N_A=N_B=k$, respectively, for $k\ge 1$.} 
\end{center}
\end{figure}

In Table~\ref{table:akbk}, we list the ten largest values of
$\alpha_k=\alpha_k^{\textrm{magic}}$ and 
$\beta_k=\alpha_k^{\textrm{deficient}}$, and in
Fig.~\ref{fig:NaNb}, we show $N_A$ and $N_B$ as 
functions of $\alpha$.
According to these results,
both $A$ and $B$ will make a single collision with the ball if $\alpha\ge 1$,
whereas if $\sqrt{2}-1=\beta_1\le \alpha<1$, the ball will hit $A$ twice and hit
$B$ only once.
As previously discussed, for any $\alpha=\alpha_k$, $N_A=N_B=k$. 
But $\alpha_k$ is the minimum value of $\alpha$ such that $N_A=k$. 
Thus, for $\alpha$ infinitesimally smaller than $\alpha_k$,
we have $N_A=k+1$ and $N_B=k$.
Similarly, note that $\beta_k$  
is the minimum value of $\alpha$ to have $N_A=N_B+1=k+1$.
Hence, for $\alpha$ infinitesimally smaller than $\beta_k$,
we have $N_A=N_B=k+1$.

\begin{figure}[!h]
\begin{center}
\includegraphics[width=6.5cm]{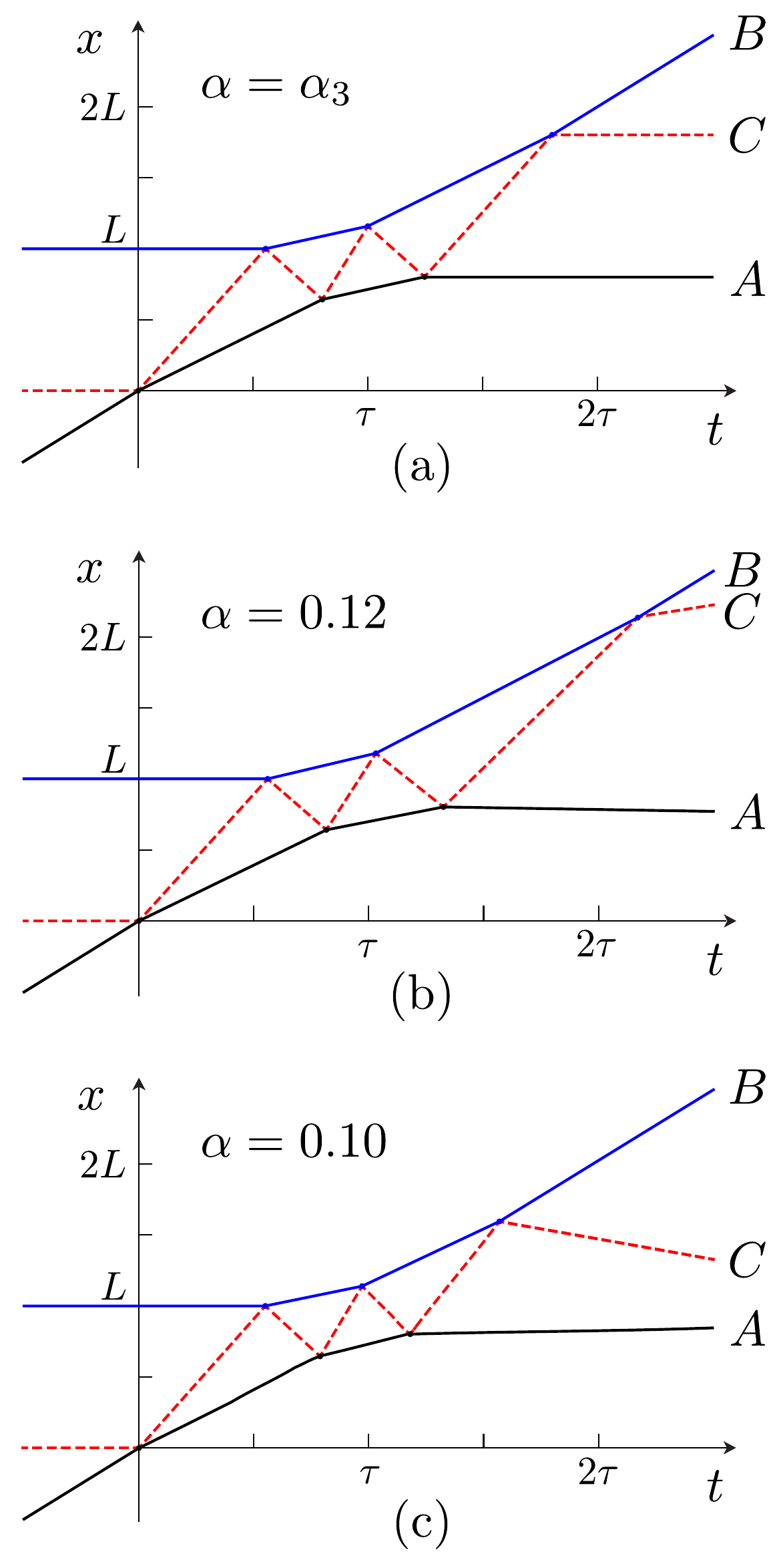}\\
\caption{\label{fig:trajectories}
The trajectories of $A$, $B$, and $C$
as functions of time
for: (a) $\alpha=\alpha_3^{\textrm{magic}}\approx 0.11$,
(b) $\alpha=0.12$, and (c)
$\alpha=0.10$, where $\tau\equiv L/V$.
The lower and upper lines represent the trajectories of
$A$ and $B$, respectively, and
the dashed line is for $C$. At $\alpha=\alpha_3^{\textrm{magic}}$ [panel (a)],
the scattered block carries the total initial momentum of the
incident block.
}
\end{center}
\end{figure}

In Fig.~\ref{fig:trajectories} we show the trajectories of $A$, $B$, and $C$
as functions of time
for (a) $\alpha=\alpha_3\approx 0.11$,
(b) $\alpha=0.12$, and (c)
$\alpha=0.10$.
When $\alpha=\alpha_3$ the trajectory of $C$ meets those of $A$ and $B$
three times each, consistent with $N_A=N_B=3$ that can be read off from
Fig.~\ref{fig:NaNb}. After $C$ hits $B$ three times, 
both $A$ and $C$ stop and the terminal velocity of $B$ is the
same as the initial velocity of the incident block.
According to Fig.~\ref{fig:NaNb},
$N_A=N_B=3$ at
$\alpha=0.12>\alpha_3$,
which is consistent with Fig.~\ref{fig:trajectories}(b). 
At this value of $\alpha$,
all three objects continue moving after their last collisions, and
although $C$ does not stop after its final collision with $B$,
the terminal speeds are such that $C$ cannot 
catch up with $B$.
At $\alpha=0.10<\alpha_3$, $N_A=4$ while $N_B=3$ according to 
Figs.~\ref{fig:NaNb} and \ref{fig:trajectories}(c).  In this scenario,
$A$ continues to move forward after its third collision with $C$;
meanwhile, $C$ moves backward after its third collision with $B$ so
that it makes another collision with $A$.

The terminal velocities of the two blocks can be
obtained by substituting $n=N_A$ and $N_B$ in 
Eqs.~(\ref{eq:NA-ans})
and (\ref{eq:NB-ans}) into the formulas for
$a_n$ and $b_n$ given in 
Eqs.~(\ref{eq:a-final}) and \eqref{eq:b-final}, to obtain
%-------------
%\begin{subequations}
%\label{eq:ab-terminal}
\begin{equation}
\label{eq:a-terminal}
a_{t}=\frac{V}{2+\alpha}
\left[1+(1+\alpha)\cos (N_A\psi)\right]
\end{equation}
and
\begin{equation}
\label{eq:b-terminal}
b_{t}=\frac{V}{2+\alpha}
\left\{1-(1+\alpha)\cos\left[(N_B+\tfrac{1}{2})\psi\right]\right\},
\end{equation}
%\end{subequations}
%-------------
where we have used $\cos(\psi/2)=(1+\alpha)^{-1}$.
By making use of momentum conservation,
we can determine the terminal velocity $c_t$ of $C$ as
%-------------
\begin{equation}
\label{eq:ct}
c_t=\frac{V-a_t-b_t}{\alpha},
\end{equation}
%-------------
which is the same as $c^{\,\prime}_{N_B}$ for 
$0<\mathfrak{s}(\pi/\psi)\le1/2$ or $c_{N_A}$ otherwise.
We can determine the terminal velocities 
at $\alpha=\alpha_k$ and $\beta_k$ by making use of Eqs.~(\ref{exact:ak})
and \eqref{exact:bk}:
when $\alpha=\alpha_k$, we have
$a_{t}=c_t=0$ and $b_{t}=V$;
when $\alpha=\beta_k$, we get
$a_{t}=-\beta_k V/(2+\beta_k)$ and
$b_t=c_t=2 V/(2+\beta_k)$.

\begin{figure}[!h]
\begin{center}
\includegraphics[width=6.5cm]{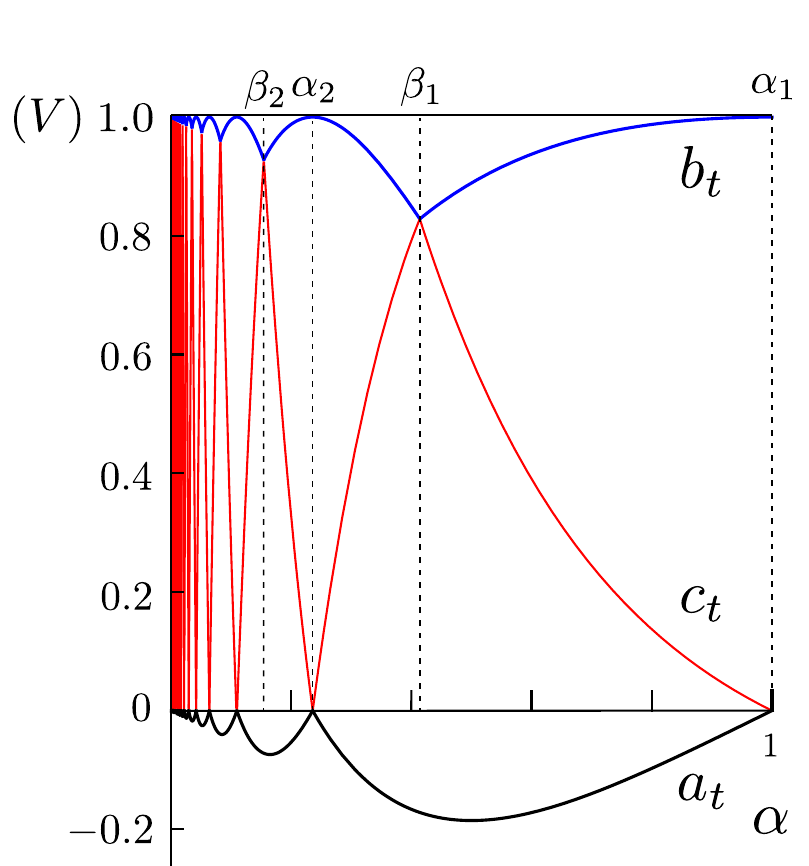}\\
\caption{\label{fig:atbtct}
The terminal velocities $a_t$, $b_t$, and $c_t$ of
$A$, $B$, and $C$, respectively, in units of the
initial velocity $V$ of $A$ as functions of the mass ratio $\alpha$.
At every $\alpha=\alpha_k$, we have
$b_t=V$ and $a_t=c_t=0$ for $k\ge 1$.
At every $\alpha=\beta_k$, we have
$b_t=c_t$, and $b_t$ becomes a local minimum for $k\ge 1$.}
\end{center}
\end{figure}

In Fig.~\ref{fig:atbtct}, we plot the terminal velocities $a_t$, $b_t$,
and $c_t$ as functions of $\alpha$.
Because the ball cannot penetrate any blocks, the three curves are ordered as
$a_t\le c_t\le b_t$ for all $\alpha$.
We see that $b_t\le V$ for all $\alpha$, as expected, because $b_t>V$
is not allowed due to energy conservation. 
According to the $a_t$ curve in Fig.~\ref{fig:atbtct}, 
the (terminal) recoil velocity of the incident block $A$ is negative\cite{footnote3} for all 
$\alpha\neq \alpha_k^{\textrm{magic}}$.

\subsection*{Inelastic collisions}

\begin{figure}[!h]
\begin{center}
\includegraphics[width=6.5cm]{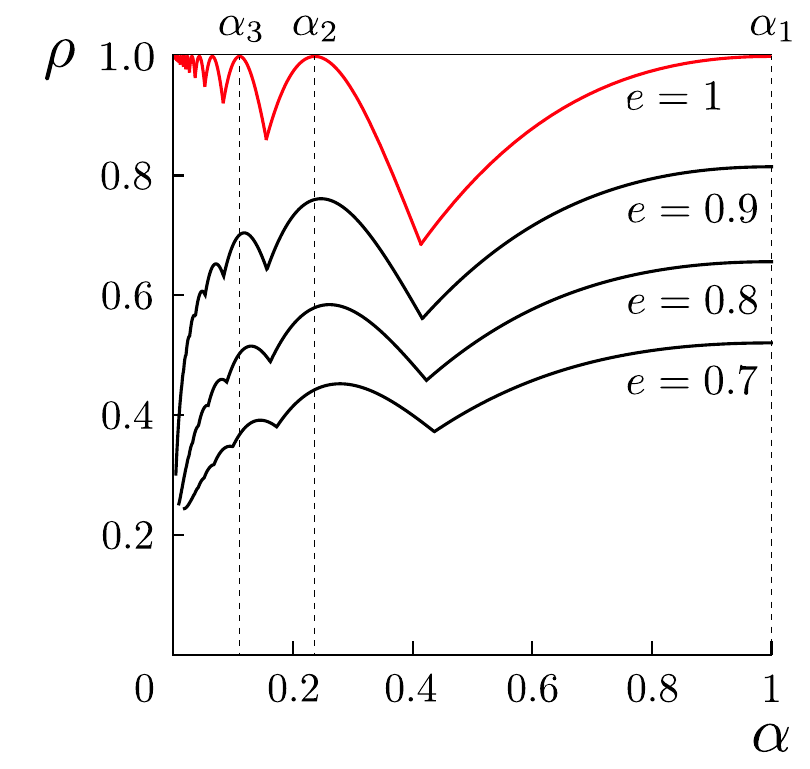}\\
\caption{\label{fig:inelastic}
The fraction
$\rho=b_t^2/V^2$ of energy transmission to the target block
as a function of $\alpha$ for different coefficients of restitution $e$.
}  
\end{center}
\end{figure}

Let us briefly investigate possible modifications of
our predictions if the ball and the blocks
do not make elastic collisions. If the collisions are not elastic,
then the energy-momentum recurrence relations 
in Eqs.~(\ref{eq:ac})--(\ref{eq:Euc-Qn})
are modified as
%-----------------------
%\begin{subequations}
%\label{eq:inelastic}
\begin{eqnarray}
%-----------------------
a_{n-1}+\alpha c^{\,\prime}_{n-1}&=&a_n+\alpha c_n,
\\
e(a_{n-1}-c^{\,\prime}_{n-1})&=&c_n-a_n,
\\
b_{n-1}+\alpha c_{n}&=&b_n+\alpha c^{\,\prime}_n
\end{eqnarray}
and
\begin{equation}
e(c_{n}-b_{n-1})=b_n-c^{\,\prime}_n,
\end{equation}
%-----------------------
%\end{subequations}
%-----------------------
where $0<e\le 1$ is the coefficient of restitution between
the ball and a block. (For an elastic collision $e=1$ and for a
completely inelastic collision $e=0$.) We follow the same procedure
as that for the elastic case to compute the
fraction of energy transmission
$\rho=b_t^2/V^2$ to the scattered block from the incident block.
In Fig.~\ref{fig:inelastic}, we show $\rho$ as a function of
$\alpha$ for $e=1$, $0.9$, $0.8$, and $0.7$.
In general we see that $\rho$ decreases
as $e$ decreases.  Furthermore, this effect is magnified as $\alpha$ decreases
because the number of collisions increases.
We also see that there are shifts to the right 
in the values of $\alpha$ where the local maxima or minima  of $\rho$ 
appear. We can compare the values of $\alpha$ with
those ($\alpha_1$, $\beta_1$, $\alpha_2$, and $\beta_2$) of the elastic case.
At $e=0.9$, for example, the shifts from
$\alpha_1$, $\beta_1$, $\alpha_2$, and $\beta_2$
are by 
$0$\,\%, $0.43$\,\%, $4.6$\,\%, $0.84$\,\%, respectively.
The size of the shift tends
to increase as $\alpha$ decreases.
For any value of $e$, the value of $\alpha$ that corresponds to
$\alpha_1=1$ of the elastic case is invariant. 

As a pedagogical example, we show a generalized version of the Newton's cradle
in Fig.~\ref{fig:system2}. There are five identical ``blocks'' $A_k$ of mass $M$
on a straight wire and four ``balls'' 
$C_k$ of mass $m_k=\alpha_k^{\textrm{magic}}M$
placed between each set of blocks.
Initially, $A_1$ is the only moving object and is traveling towards $C_1$.
We assume that all collisions are elastic and neglect friction.
Any two adjacent blocks $A_k$ and $A_{k+1}$ exchange
momenta through (possibly multiple) collisions with
$C_k$ while the remainder of the system is completely at rest. 
After $A_k$ makes the $k$th collision with $C_k$, $A_k$ stops and
$C_k$ makes the $k$th collision with $A_{k+1}$.
At this point $C_k$ stops and
$A_{k+1}$ carries the entire incident momentum of $A_{k}$.
This process continues until $A_5$ carries the initial momentum of $A_1$.
After that, $A_5$ bounces back against the wall to continue a similar set of
multiple collisions in the opposite direction.
This experiment can be
carried out by making use of an air track that is typically available
in undergraduate physics laboratories.

%%%%%%%%%%%%%%%%%%%%%%%%%%%%%%%%%%%%%%%%
\section{\label{sec:concl}Conclusion}
%%%%%%%%%%%%%%%%%%%%%%%%%%%%%%%%%%%%%%%%

We have considered the one-dimensional scattering of two identical blocks
of mass $M$ that exchange energy and momentum through elastic collisions
with an intermediary ball of mass $m=\alpha M$.
Because the ball and the target block are initially not in contact with each other,
the system experiences multiple two-body collisions that are well ordered
and therefore the motion of the system is  uniquely determined by
energy and momentum conservation.

A vector sequence $\textbf{V}_n$ was constructed by rescaling 
the velocities of the three objects in each cycle of multiple collisions.
We then showed that the energy-momentum conservation laws
result in transforming the vector $\textbf{V}_n$ through a
(pure) rotation: $\textbf{V}_n=\Lambda^n\textbf{V}_{0}$, with
$\textrm{det}[\Lambda]=+1$ and $\textbf{V}_{n}^2=\textbf{V}_{0}^2=V^2$.
Recursive use of the covariant recurrence relation
leads to analytic expressions for the velocities of the three objects after each
collision. Based on these results, we have computed the total number of collisions
and the terminal velocity for each object in terms of $\alpha$.
 
The covariant approach was shown to be equivalent to
the billiard-ball mapping approach. While the billiard-ball mapping
approach reformulates the 
problem into a problem of projective geometry,
the covariant approach employs elementary techniques of group theory and
differential geometry that are well described in standard textbooks on
classical mechanics or mathematical physics. We expect that the covariant
approach presented in this work can 
be systematically generalized to the relativistic case.

It is rather remarkable that at magic mass ratios $\alpha=\alpha_k^{\textrm{magic}}$,
the energy and momentum of the incident block are completely
transferred to the target block. Such complete momentum transfer to a single
scattered particle in many-body collisions is difficult to find. The only nontrivial
example is Newton's cradle, which is equivalent to our model when $\alpha=1$.
As a pedagogical example, we have devised a generalized version of Newton's
cradle (Fig.~\ref{fig:system2}) that can be tested in an air track experiment.

We have also verified the identities
$\alpha_k^{\textrm{magic}}=\alpha_k$ and $\alpha_k^{\textrm{deficient}}=\beta_k$.
In inelastic collisions, 
the peaks of
the energy transfer fraction $\rho$ shift from
$\alpha_k^{\textrm{magic}}$ and the effect increases as $\alpha$ decreases.

%%%%%%%%%%%%%%%%%%%%%%%%%%%%%%%%%%%%%%%%
\begin{acknowledgments}
J.L.\ wishes to express his gratitude to the High Energy Theory Group at
Argonne National Laboratory, where part of his work was carried out 
on a leave of absence from Korea University, for its hospitality.
We thank U-Rae Kim for providing useful information
on the package \textit{animate} that enabled us to generate the animation shown 
in Fig.~\ref{fig:system2}.
This work was supported in part by Korea University.
The work of J.-H.E.\ is supported by a Global Ph.D.\ Fellowship Program
through the National Research Foundation of Korea  funded 
by the Ministry of Education under Contract
No.\ NRF-2012H1A2A1003138.
\end{acknowledgments}
%%%%%%%%%%%%%%%%%%%%%%%%%%%%%%%%%%%%%%%%

%%%%%%%%%%%%%%%%%%%%%%%%%%%%%%%%%%%%%%%%%%%%%%%%%%%%%
\appendix
%--------------------------------------------------
\section{Orthogonal matrices\label{app:orthogonal}}
%--------------------------------------------------
Here, we list useful properties
of orthogonal matrices in two and three dimensions
that are frequently used in the text.
Detailed proofs and examples can be found
in standard textbooks on classical mechanics or mathematical physics,
such as Refs.~\citenum{Marion} and \citenum{Arfken}.

\begin{itemize}
\item
If there is a linear transformation $\textbf{V}\to\textbf{V}'=\mathcal{O} \textbf{V}$ of
a Euclidean vector $\textbf{V}$ that preserves its magnitude
($\textbf{V}^{\prime 2}=\textbf{V}^2$), then
the transformation matrix $\mathcal{O}$ must be \textit{orthogonal}; the inverse of 
$\mathcal{O}$ is identical to
its transpose: $\mathcal{O}^{-1}=\mathcal{O}^T$.
\item
The determinant of an orthogonal matrix $\mathcal{O}$ is either
$+1$ or $-1$: $\textrm{det}[\mathcal{O}]=\pm 1$.
\item
A linear transformation $\textbf{V}'=\mathbbm{P}_k\textbf{V}$
is called  a \textit{reflection} if
$V'_i=V_i$ for all $i\neq k$ and $V'_k=-V_k$.
The magnitude of a vector is invariant under reflection.
\item
In a two-dimensional Euclidean space, any orthogonal matrix 
$\mathcal{O}$ with $\textrm{det}[\mathcal{O}]=+1$ can be parametrized by $\mathcal{O}=\lambda(\phi)$, where
$\lambda(\phi)$ is the $2\times 2$ rotation matrix by angle $\phi$:
\begin{equation}
\lambda(\phi)=
\begin{pmatrix}
\cos\phi&-\sin\phi\\
\sin\phi&\cos\phi
\end{pmatrix}.
\end{equation}
If $\textrm{det}[\mathcal{O}]=-1$, then 
$\mathcal{O}$ can be parametrized by $\mathcal{O}=\lambda(\phi')\mathbbm{P}_2$,
which is the product of a rotation by angle $\phi'$ and a reflection
$\mathbbm{P}_2=\textrm{diag}[1,-1]$. Here, $\textrm{diag}[a_1,a_2,\cdots,a_n]$
stands for the diagonal matrix $A$ whose diagonal elements are $A_{ii}=a_i$.
\item
In three dimensions, 
any orthogonal matrix
$\mathcal{O}$ with $\textrm{det}[\mathcal{O}]=+1$ can be parametrized by 
$\mathcal{O}=\lambda_{\hat{n}}(\phi)$, where 
$\lambda_{\hat{n}}(\phi)$ is the rotation matrix about the axis parallel to 
a unit vector
$\hat{n}$ by angle $\phi$. If $\hat{n}$ is 
a unit basis vector $\hat{e}_i$ of a Cartesian coordinate system $S$,
then we write $\lambda_{\hat{n}}(\phi)=\lambda_{i}(\phi)$, where
%\begin{subequations}
%\label{eq:lambda-i}
\begin{equation}
\label{eq:lambda-1}
\lambda_1(\phi)=
\begin{pmatrix}
1&0&0\\
0&\phantom{,}\cos\phi&-\sin\phi\\
0&\phantom{,}\sin\phi&\phantom{-}\cos\phi
\end{pmatrix},
\end{equation}
\begin{equation}
\label{eq:lambda-2}
\lambda_2(\phi)=
\begin{pmatrix}
\phantom{,}\cos\phi&\phantom{,}0\phantom{,}&\sin\phi\\
0&\phantom{,}1\phantom{,}&0\\
-\sin\phi&\phantom{,}0\phantom{,}&\cos\phi
\end{pmatrix},
\end{equation}
and
\begin{equation}
\label{eq:lambda-3}
\lambda_3(\phi)=
\begin{pmatrix}
\cos\phi&-\sin\phi&\phantom{,}0\\
\sin\phi&\phantom{-}\cos\phi&\phantom{,}0\\
0&0&\phantom{,}1
\end{pmatrix}.
\end{equation}
%\end{subequations}
\item
If the axis of rotation $\hat{n}$ is not parallel to any axis
of a Cartesian coordinate system $S$,
then we can choose a new coordinate system $S'$ whose triad is
$\{\hat{e}_1',\hat{e}'_2,\hat{e}'_3=\hat{n}\}$.
Note that the triple scalar product of the three basis vectors of a triad is $\hat{e}_i\cdot\hat{e}_j\times\hat{e}_k=\epsilon_{ijk}$, where
$\epsilon_{ijk}$ is the Levi-Civita tensor that is antisymmetric
under exchange of any two indices and $\epsilon_{123}=+1$.
Because the matrix representation of 
$\lambda_{\hat{n}}(\phi)$ in $S'$ is $\lambda_3(\phi)$, 
that in $S$ can be expressed as
\begin{equation}
\label{eq:lambda-n}
\lambda_{\hat{n}}(\phi)=
R\,\lambda_3(\phi)R^T,
\end{equation}
where the matrix elements of $R$ in $S$ are given by
\begin{equation}
\label{def:R}
R_{ij}=\hat{e}_i\cdot \hat{e}_j^\prime.
\end{equation}
It is trivial to show that $R$ is orthogonal.
If $\textrm{det}[\mathcal{O}]=-1$, then 
$\mathcal{O}$ is parametrized by $\mathcal{O}=\lambda_{\hat{n}'}(\phi')\mathbbm{P}_3$,
which is the product of a rotation
about $\hat{n}'$ by angle $\phi'$ and a reflection
$\mathbbm{P}_3=\textrm{diag}[1,1,-1]$.
\end{itemize}

%===============================================================
\section{\label{app:Vn}Derivation of Eq.~(\ref{eq:V-n})}
%===============================================================

Here, we derive the recurrence relation in
 Eq.~(\ref{eq:V-n}).
The constraints in Eqs.~(\ref{eq:ac})--(\ref{eq:Euc-Qn})
due to energy-momentum conservation result in
%-----------------------
%\begin{subequations}
%\label{eq:recur2d}
\begin{equation}
\label{eq:abc-n}
\begin{pmatrix}
a_n\\\sqrt{\alpha}c_n
\end{pmatrix}
=\Gamma
\begin{pmatrix}
a_{n-1}
\\
\sqrt{\alpha}c^{\,\prime}_{n-1}
\end{pmatrix},
\end{equation}
and
\begin{equation}
\label{eq:abcp-n}
\begin{pmatrix}
b_n\\\sqrt{\alpha}c^{\,\prime}_n
\end{pmatrix}
=\Gamma
\begin{pmatrix}
b_{n-1}
\\
\sqrt{\alpha}c_{n}
\end{pmatrix},
\end{equation}
%\end{subequations}
%-----------------------
where the $2\times 2$ matrix $\Gamma$
is defined by
%-----------------------
\begin{equation}
\label{eq:Gamma}
%-----------------------
\Gamma=\frac{1}{1+\alpha}\begin{pmatrix}
1-\alpha&2\sqrt{\alpha}\\
2\sqrt{\alpha}&-(1-\alpha)
\end{pmatrix}.
%-----------------------
\end{equation}
%-----------------------
According to Eqs.~(\ref{eq:Euc-Pn}) and (\ref{eq:Euc-Qn}), 
the magnitudes of the column vectors in each of 
Eqs.~(\ref{eq:abc-n}) and (\ref{eq:abcp-n})
are invariant. Therefore,
$\Gamma$ can be parametrized by the product of
a rotation matrix $\lambda(\theta)$ and a reflection matrix $\mathbbm{P}_{2}$ as
%-----------------------
\begin{equation}
\label{def:gamma}
\Gamma=\begin{pmatrix}
\cos\theta&\sin\theta\\
\sin\theta&-\cos\theta
\end{pmatrix}=\lambda(\theta)\mathbbm{P}_{2}.
\end{equation}
%-----------------------
Because $\textrm{det}[\lambda(\theta)]=1$ and
$\textrm{det}[\mathbbm{P}_{2}]=-1$, we have
$\textrm{det}[\Gamma]=
\textrm{det}[\lambda(\theta)]\textrm{det}[\mathbbm{P}_{2}]=-1$. 
Here, the parameter $\theta$ is given by
%-----------------------
\begin{equation}
%-----------------------
\label{def:theta}
\theta=2\arctan\sqrt{\alpha}.
%-----------------------
\end{equation}
%-----------------------
Because we restrict ourselves to the case $0<\alpha<1$,
the range of $\theta$ is $0< \theta<\pi/2$.~\cite{footnote2}
(For small
$\alpha$, $\theta\approx 2\sqrt{\alpha}$.) In Table~\ref{table:trig},
we list the values for trigonometric functions 
at $\theta$ and $\theta/2$.

By making use of Eqs.~(\ref{eq:abc-n})--(\ref{def:gamma}),
we can find the recurrence relation for the 
three-dimensional Euclidean-vector sequence $\textbf{V}_n$ of the form
%------------------
\begin{equation}
\label{eq:V-n-app}
\textbf{V}_n=\Lambda \textbf{V}_{n-1}.
\end{equation}
%------------------
Then $\textbf{V}_n$ can be expressed in terms of the initial value 
$\textbf{V}_0=(V\,0\,\,0)^T$ as
%------------------
\begin{equation}
\label{eq:V-n-0-app}
\textbf{V}_n=\Lambda^n \textbf{V}_0.
\end{equation}
%------------------

Next, we find the matrix representation of
the matrix $\Lambda$.
We first verify that
$\Lambda$ is
 a $3\times3$ matrix for a pure rotation.
We combine the two relations in 
Eqs.~(\ref{eq:abc-n}) and (\ref{eq:abcp-n})
to express $\Lambda$ as
\begin{equation}
\label{Lam-GAM}
\Lambda=\Gamma_B\Gamma_A,
\end{equation}
where the matrices $\Gamma_A$ and $\Gamma_B$  
are the $3\times 3$ generalizations of $\Gamma$
in Eqs.~(\ref{eq:abc-n}) and (\ref{eq:abcp-n}):
%------------------
%\begin{subequations}
%\label{GA-GB}
\begin{equation}
\Gamma_A=
\lambda_2(-\theta)\mathbbm{P}_3=
\begin{pmatrix}
\cos\theta&\,\,\,0&\phantom{-}\sin\theta\\
0&\,\,\,1&\,0\\
\sin\theta&\,\,\,0&-\cos\theta
\end{pmatrix}
\end{equation}
and
\begin{equation}
\Gamma_B=
\lambda_1(\theta)\mathbbm{P}_3\phantom{-}=
\begin{pmatrix}
\,\,1\,\,&0&\,0\\
\,\,0\,\,&\cos\theta&\phantom{-}\sin\theta\\
\,\,0\,\,&\sin\theta&-\cos\theta
\end{pmatrix}.
\end{equation}
%\end{subequations}
%------------------
Here, $\theta$ and $\lambda_i(\theta)$ are defined in Eqs.~(\ref{def:theta}) and 
(\ref{eq:lambda-1})--(\ref{eq:lambda-3}), respectively, and
$\mathbbm{P}_3=\textrm{diag}[1,1,-1]$.
The explicit form of the matrix $\Lambda$ is
%---------------------
\begin{equation}
\label{eq:Lambda-app}
%---------------------
\Lambda=
\begin{pmatrix}
\cos\theta&0&\sin\theta\\
\sin^2\theta&\cos\theta&-\sin\theta\cos\theta\\
-\sin\theta\cos\theta&\sin\theta& \cos^2\theta
\end{pmatrix},
%=
%\begin{pmatrix}
%\frac{1-\alpha}{1+\alpha}&\,\,\,\,\,
%0\,\,\,\,&
%\frac{2\sqrt{\alpha}}{1+\alpha}
%\\
%\frac{4\alpha}{(1+\alpha)^2}&\,\,\,\,\,
%\frac{1-\alpha}{1+\alpha}\,\,\,\,&
%-\frac{2(1-\alpha)\sqrt{\alpha}}{(1+\alpha)^2}
%\\
%-\frac{2(1-\alpha)\sqrt{\alpha}}{(1+\alpha)^2}&\,\,\,\,\,
%\frac{2\sqrt\alpha}{1+\alpha}\,\,\,\,&
%\frac{(1-\alpha)^2}{(1+\alpha)^2}
%\end{pmatrix}.
%---------------------
\end{equation}
%---------------------
which, using Table~\ref{table:trig}, results in Eq.~(\ref{eq:Lambda}).

Because $\textrm{det}[\lambda_i(\theta)]=1$ and
$\textrm{det}[\mathbbm{P}_3]=-1$, we know that
$\textrm{det}[\Gamma_A]=\textrm{det}[\Gamma_B]=-1$ 
and $\textrm{det}[\Lambda]=1$. In addition, direct computation
shows that $\Lambda$ is orthogonal:
$\Lambda^{-1}=\Lambda^T$.
Therefore, the matrix $\Lambda$ represents a pure rotation.

%===============================================================
\section{Computation of $\bm{\Lambda^n}$\label{app:lam-n}}
%===============================================================

Here, we derive the analytic expression for
the matrix $\Lambda^n$ that is necessary to compute $\textbf{V}_n$
in Eq.~(\ref{eq:V-n-0}). Let $\hat{n}$ and $\psi$ be the axis and
angle of rotation for the matrix $\Lambda$, respectively. 
If we choose a 
Cartesian coordinate system $S'$, in which $\hat{n}=\hat{e}_3^\prime$, 
the matrix representation of $\Lambda$ is of the form
%---------------------
\begin{equation}
\label{eq-app:lambda-R}
%---------------------
\Lambda^\prime=\lambda_3(\psi).
%---------------------
\end{equation}
%---------------------
In $S'$, $\Lambda^n$ can be expressed as a single rotation:
%---------------------
\begin{equation}
%---------------------
(\Lambda^\prime)^n=[\lambda_3(\psi)]^n=\lambda_3(n\psi).
%---------------------
\end{equation}
%---------------------
Then the matrices $\Lambda$ and $\Lambda^n$ defined in a general
frame $S$ can be expressed as
%---------------------
%\begin{subequations}
\begin{equation}
%---------------------
\Lambda=R\lambda_3(\psi)R^T
\end{equation}
and
\begin{equation}
\label{eq:Lambda-n-final} 
\Lambda^n=R\lambda_3(n\psi)R^T,
%---------------------
\end{equation}
%\end{subequations}
%---------------------
where the coordinate transformation matrix $R$
is defined  in Eq.~(\ref{def:R}).

By generalizing the method in Ref.~\citenum{Ee:2012} for two dimensions
into three dimensions, we determine the transformation matrix $R$.
Because $\hat{e}'_3$ is invariant under rotation $\Lambda$, we have
%---------------------
\begin{equation}
\Lambda\hat{e}'_3=\hat{e}'_3.
\end{equation}
%---------------------
The solution for this constraint equation is 
$\hat{e}_3^{\prime}=(1\; 1\;\sqrt{\alpha})^T/\sqrt{2+\alpha}$.
Choosing the remaining two bases for the coordinate system $S'$
as $\hat{e}_1^{\prime}=
\sqrt{1+(\alpha/2)}\;\hat{e}_3\times\hat{e}_3^{\prime}$ and 
$\hat{e}_2^{\prime}=\hat{e}_3^{\prime}\times \hat{e}_1^{\prime}$,
we determine the triad 
$\{\hat{e}_1^{\prime},\hat{e}_2^{\prime},\hat{e}_3^{\prime}\}$
of $S'$. As a result, we find that
%---------------------
\begin{equation}
\label{eq:R-app}
%---------------------
R=(\hat{e}_1^{\prime}\,\,\hat{e}_2^{\prime}\,\,\hat{e}_3^{\prime})=
\begin{pmatrix}
\displaystyle{-\frac{1}{\sqrt{2}}}&
\displaystyle{-\sqrt{\frac{\alpha}{2(2+\alpha)}}}&
\displaystyle{\frac{1}{\sqrt{2+\alpha}}}
\\[10 pt]
\displaystyle{\frac{1}{\sqrt{2}}}&
\displaystyle{-\sqrt{\frac{\alpha}{2(2+\alpha)}}}&
\displaystyle{\frac{1}{\sqrt{2+\alpha}}}
\\[10 pt]
0&
\displaystyle{\sqrt{\frac{2}{2+\alpha}}}&
\displaystyle{\frac{\sqrt{\alpha}}{\sqrt{2+\alpha}}}
\end{pmatrix},
%---------------------
\end{equation}
%---------------------
and the parameter $\psi$ is defined by
%-------------
\begin{equation}
\label{def:psi-app}
\psi\equiv 2\arctan\sqrt{\alpha(2+\alpha)}.
\end{equation}
%-------------
For $0< \alpha<1$,
the range of $\psi$ is $0< \psi<2\pi/3$ (for small
$\alpha$, $\psi\approx 2\sqrt{2\alpha}$).
In Table \ref{table:trig}, we list
the values for trigonometric functions 
at $\psi$, $\psi/2$, and $\psi/4$.

%===============================================================
\section{Special values for trigonometric functions\label{sec:tri}}
%===============================================================

Here, we summarize a way to evaluate 
the trigonometric functions at special values such as 
$n\theta$, $(n+\tfrac{1}{2})\theta$, $n\psi$, $(n+\tfrac{1}{2})\psi$,
and $(n+\tfrac{1}{4})\psi$ in terms of $\alpha$. The parameters 
$\theta$ and $\psi$ are defined in Eqs.~(\ref{theta-txt})
and (\ref{def:psi}), respectively.

%===============================================================
\subsection{At angles $\bm{nx}$\label{sec:nx}}
%===============================================================

We can compute
$\cos nx$ and $\sin nx$ 
for $x=\theta$, $\psi$ as
%-------------
%\begin{subequations}
\begin{equation}
\cos nx=
\mathfrak{Re}[e^{inx}]
=\sum_{k=0}^{\lfloor n/2\rfloor}
\frac{(-1)^k n!}{(2k)!(n-2k)!}\sin^{2k} x \cos^{n-2k}x
\end{equation}
and
\begin{equation}
\sin nx=
\mathfrak{Im}[e^{inx}]
=\sum_{k=0}^{\lfloor(n-1)/2\rfloor}
\frac{(-1)^k n!}{(2k+1)!(n-2k-1)!}\sin^{2k+1} x \cos^{n-2k-1}x,
\end{equation}
%\end{subequations}
%-------------
where the floor function $\lfloor x\rfloor$ is defined in
Eq.~(\ref{def:floor}). Special values for
$\cos x$ and $\sin x$ for $\theta$ and $\psi$ are given in
Table~\ref{table:trig}.

%================================================================
\subsection{At angles 
$\bm{(n+\frac{1}{4})x}$ and
$\bm{(n+\frac{1}{2})x}$\label{sec:nhx}}
%================================================================

We can compute
$\cos (n+r)x$ and $\sin (n+r)x$ 
for $x=\theta$, $\psi$ and
$r=1/2$, $1/4$ as
%-------------
%\begin{subequations}
\begin{equation}
\cos (n+r)x=
\cos nx \cos r x-
\sin nx \sin r x
\end{equation}
and
\begin{equation}
\sin (n+r)x=
\sin nx \cos r x+
\cos nx \sin r x.
\end{equation}
%\end{subequations}
%-------------
The values for
$\cos (\theta/2)$,
$\sin (\theta/2)$,
$\cos (\psi/2)$,
$\sin (\psi/2)$,
$\cos (\psi/4)$, and
$\sin (\psi/4)$ are given in
Table~\ref{table:trig}.

%===============================================================
\section{Computation of $\bm{N_A}$ and $\bm{N_B}$\label{sec:N-app}}
%===============================================================

Here, we compute the total number of collisions $N_i$
between the ball and the block $i$ for $i=A$ or $B$.  In computing $N_A$ and $N_B$,
it is convenient to use 
the ceiling ($\lceil x\rceil$) and floor 
($\lfloor x\rfloor$) functions. 
For any real number $x$ they are defined by
%-------------
%\begin{subequations}
\begin{equation}
\label{def:ceil}
\lceil x\rceil=n\;\;\textrm{satisfying}\;\; n-1< x\le n
\end{equation}
and
\begin{equation}
\label{def:floor}
\lfloor x\rfloor=n\;\;\textrm{satisfying}\;\; n\le x<n+1,
\end{equation}
%\end{subequations}
%-------------
where $n$ is a unique integer in each case. In addition,
the sawtooth function $\mathfrak{s}(x)$ is defined by
%-------------
\begin{equation}
\label{def:sawtooth}
\mathfrak{s}(x)=x-\lfloor x\rfloor,
\end{equation}
%-------------
where the range is $0\le \mathfrak{s}(x) <1$.

The collision at $P_n$ is allowed only if $a_{n-1}>c_{n-1}^{\,\prime}$.
After the collision at $P_n$, the velocities of $A$ and $C$
become $a_n$ and $c_n$, respectively.
If the ball $C$ experiences another collision with $B$,
then it returns to $A$ with the velocity $c^{\,\prime}_n$.
If $n=N_A$, then
$A$'s velocity right after $P_n$ must be smaller than or equal 
to that of $C$ right after $Q_n$, or
%-------------
\begin{equation}
\label{eq:NA-constraint}
a_{N_A} \le c^{\,\prime}_{N_A}.
\end{equation}
%-------------
By making use of Eqs.~(\ref{eq:a-final})--\eqref{eq:c'-final}, 
(\ref{eq:NA-constraint}),
and Table~\ref{table:trig},
we find that 
%-------------
\begin{equation}
\label{eq:NA-constraint-simple}
\sin\left[\left(N_A-\tfrac{1}{2}\right)\psi\right]> 0\,\,\textrm{ and }\,\,
\sin\left[\left(N_A+\tfrac{1}{2}\right)\psi\right]\leq 0,
\end{equation}
%-------------
with the solution of
%-------------
\begin{equation}
\label{eq:NA-ans}
N_{A}=\left\lceil \frac{\pi}{\psi}-\frac{1}{2}\right\rceil.
\end{equation}
%-------------

In a similar manner, we can find the constraint on $N_B$.
The collision at $Q_n$ is allowed only if $c_{n}>b_{n-1}$.
After the collision with $B$ at $Q_n$, $C$ makes another collision
with $A$ at $P_{n+1}$, resulting in the velocity $c_{n+1}$. 
Therefore, if $n=N_B$, we then have
%-------------
\begin{equation}
\label{eq:NB-constraint}
c_{N_B+1} \leq  b_{N_B},
\end{equation}
%-------------
which leads to
%-------------
\begin{equation}
\label{eq:NB-constraint-simple}
\sin \left(N_B\psi\right)>0\quad\textrm{ and }\quad\sin\left[\left(N_B+1\right)\psi\right]\leq0,
\end{equation}
%-------------
with the solution of
%-------------
\begin{equation}
\label{eq:NB-ans}
N_{B}=\left\lceil \frac{\pi}{\psi}-1\right\rceil.
\end{equation}
%-------------

According to Eqs.~(\ref{eq:NA-ans}) and (\ref{eq:NB-ans}), we find that
%-------------
\begin{equation}
\label{eq:NANB-ans}
N_A
=
\begin{cases}
N_B=\lceil\pi/\psi\rceil-1&
{\rm for}~~0<\mathfrak{s}(\pi/\psi)\le 1/2,
\\
N_B+1=\lceil\pi/\psi\rceil,&
\textrm{otherwise}.
\end{cases}
\end{equation}
%-------------
If $\alpha$ is small, then we can approximate Eq.~(\ref{def:psi})
as $\psi\approx 2\sqrt{2\alpha}$ to obtain
%-------------
\begin{equation}
\label{eq:NA-approx}
N_{A}\approx N_{B}\approx 
\frac{\pi}{2\sqrt{2\alpha}},
\end{equation}
%-------------
which is consistent with a previous 
result given in Eq.~(6) of Ref.~\citenum{Redner:2004}, where the author
counts the number of collisions on both sides of the ball.

%===============================================================
\section{Proof of $\bm{\alpha_k^{\textrm{magic}}=\alpha_k}$\label{app:magic}}
%===============================================================

Here, we verify that $\alpha_k$ in Eq.~(\ref{exact:ak})
is the $k$th magic mass ratio $\alpha_k^{\textrm{magic}}$ at which 
$a_t=c_t=0$ and $b_t=V$.

Verification of the largest magic mass ratio 
$\alpha_1^{\textrm{magic}}=1$ is trivial.
In general, energy and momentum conservation requires that the 
condition for the magic mass ratio $a_t=c_t=0$ and $b_t=V$ can be 
reduced into the single condition $b_t=V$. Applying this requirement
to Eq.~(\ref{eq:b-terminal}), we find that
%-------------
\begin{equation}
\label{eq:magic-b-cond}
\cos[(N_B+\tfrac{1}{2})\psi]=-1.
\end{equation}
%-------------
Because the contact force on the block $B$ due to the collision with $C$
is always along the positive $x$-axis, the acceleration of $B$ is never 
negative. Therefore, $b_t$ in Eq.~(\ref{eq:b-terminal}) is never decreasing,
which requires that
%-------------
\begin{equation}
\label{eq:magic-b-cond-simple}
N_B
=
\frac{\pi}{\psi}-\frac{1}{2}.
\end{equation}
%-------------
If we substitute the value for $\psi$ into Eq.~(\ref{def:psi}),
then the constraint is equivalent to a quadratic equation,
whose unique solution is given by Eq.~(\ref{exact:ak}).

%===============================================================
\section{Proof of $\bm{\alpha_k^{\textrm{deficient}}=\beta_k}$\label{app:deficient}}
%===============================================================

In this appendix, we verify that $\beta_k\!$ defined in 
Eq.~(\ref{exact:bk}) is identical to the deficient mass ratio $\alpha_k^{\textrm{deficient}}$
at which $b_t$ reaches the local minimum within the region 
$\alpha_{k+1}<\alpha<\alpha_k$.

As shown in Fig.~\ref{fig:atbtct}, the terminal velocity $b_t$ has
local minima. At each local minimum,
the ball is at a local maximum and
$b_t=c_t$.
Next, we verify the statement 
that $b_t=c_t$ at $\alpha=\beta_k$ 
for any $k\ge 1$. 
If we require
$b_t=c_t$, then conservation of energy and momentum lead to
%-------------
%\begin{subequations}
\begin{equation}
(1+\alpha)b_t+a_t=V
\end{equation}
and
\begin{equation}
(1+\alpha)b^2_t+a^2_t=V^2.
\end{equation}
%\end{subequations}
%-------------
These coupled quadratic equations have a trivial solution
$b_t=0$ and $a_t=V$ that is equivalent to the initial condition.
Because the three objects do not penetrate each other, we discard
this trivial solution. Then we have a unique set of solutions:
%-------------
%\begin{subequations}
\begin{equation}
a_t=-\frac{\alpha V}{2+\alpha}
\end{equation}
and
\begin{equation}
\label{eq:defb-terminal}
b_t=\frac{2 V}{2+\alpha}. 
\end{equation}
%\end{subequations}
%-------------
By comparing Eq.~(\ref{eq:b-terminal}) and Eq.~(\ref{eq:defb-terminal}),
we find that 
%-------------
\begin{equation}
\label{eq:def-b-cond}
\cos[(N_B+\tfrac{1}{2})\psi]=-\frac{1}{1+\alpha},
\end{equation}
%-------------
with a solution of
%-------------
\begin{equation}
N_B=\frac{\pi}{\psi}-1,
\end{equation}
%-------------
which is equivalent to Eq.~(\ref{exact:bk})
with the value for $\psi$ in Eq.~(\ref{def:psi}).

%%%%%%%%%%%%%%%%%%%%%%%%%%%%%%%%%%%%%%%%%%%%%%%%%%%%%

\newpage
\section*{Tables}
%%%%%%%%%%%%%%%%%%%%%%%%%%%%%%%%%%%%%%%%%%%%%%%%%%%%%
%==============================================================
\begin{table}[!h]
\begin{center}
\caption{\label{table:trig}
The values of $\cos x$, $\sin x$, and $\tan x$ at
$x=\theta/2$, $\theta$, $\psi/4$,
$\psi/2$, and $\psi$ as functions of $\alpha$,
where $\theta=2\arctan\sqrt{\alpha}$ and
$\psi=2\arctan\sqrt{\alpha(2+\alpha)}$.
}
\vspace{10 pt}
\begin{ruledtabular}
\begin{tabular}{cccc} 
%---------------------------------------------------
$\displaystyle x\,\backslash\,f(x)$ & 
$\displaystyle \cos x$ \phantom{xx}&
$\displaystyle \sin x$ \phantom{xx}& 
$\displaystyle \tan x$ \phantom{xx}\\[1ex]
\hline
&&&\\[-3.5ex]
%---------------------------------------------------
$\displaystyle \frac{\theta}{2}$ & 
$\displaystyle \frac{1}{\sqrt{1+\alpha}}$ & 
$\displaystyle \sqrt{\frac{\alpha}{1+\alpha}}$ & 
$\displaystyle \sqrt{\alpha}$ \\[3ex]
%---------------------------------------------------
$\displaystyle \theta$&
$\displaystyle \frac{1-\alpha}{1+\alpha}$&
$\displaystyle \frac{2\sqrt{\alpha}}{1+\alpha}$ &
$\displaystyle \frac{2\sqrt{\alpha}}{1-\alpha}$ \\[3ex]
%---------------------------------------------------
$\displaystyle \frac{\psi}{4}$ & 
$\displaystyle \sqrt{\frac{2+\alpha}{2(1+\alpha)}}$ & 
$\displaystyle \sqrt{\frac{\alpha}{2(1+\alpha)}}$ & 
$\displaystyle \sqrt{\frac{\alpha}{2+\alpha}}$ \\[3ex]
%---------------------------------------------------
$\displaystyle \frac{\psi}{2}$ & 
$\displaystyle \frac{1}{1+\alpha}$ &
$\displaystyle \frac{\sqrt{\alpha(2+\alpha)}}{1+\alpha}$ &
$\displaystyle \sqrt{\alpha(2+\alpha)}$ \\[3ex]
%---------------------------------------------------
$\displaystyle \psi$ &
$\displaystyle 
\frac{1-2\alpha-\alpha^2}{(1+\alpha)^2}$ &
$\displaystyle 
\frac{2\sqrt{\alpha(2+\alpha)}}{(1+\alpha)^2}$&
$\displaystyle 
\frac{2\sqrt{\alpha(2+\alpha)}}{1-2\alpha-\alpha^2}$ 
\\[1ex]
\end{tabular}
\end{ruledtabular}
\end{center}
\end{table}
%==============================================================
%=====================================================================
\begin{table}[!h]
\begin{center}
\caption{
The ten largest values for the magic mass ratio
$\alpha_k^{\textrm{magic}}$ and the deficient mass ratio
$\alpha_k^{\textrm{deficient}}$
that are defined in Eqs.~(\ref{exact:ak}) and \eqref{exact:bk}.
The proofs of the relations with threshold mass ratios
$\alpha_k^{\textrm{magic}}=\alpha_k$
and
$\alpha_k^{\textrm{deficient}}=\beta_k$
are given in Appendices \ref{app:magic} and \ref{app:deficient}, respectively.
\label{table:akbk}}
\vspace{10 pt}
\begin{ruledtabular}
\begin{tabular}{ccc}
$k$&$\alpha_k^{\textrm{magic}}=\alpha_k$&
$\alpha_k^{\textrm{deficient}}=\beta_k$\\
\hline
$1$& $1$ & $0.414214$\\
$2$& $0.236068$ & $0.154701$\\
$3$& $0.109916$ & $0.082392$\\
$4$& $0.064178$ & $0.051462$\\
$5$& $0.042217$ & $0.035276$\\
$6$& $0.029927$ & $0.025717$\\
$7$& $0.022341$ & $0.019591$\\
$8$& $0.017321$ & $0.015427$\\
$9$& $0.013827$ & $0.012465$\\
$10$& $0.011295$& $0.010283$
\end{tabular}
\end{ruledtabular}
\end{center}
\end{table}
%=====================================================================

%%%%%%%%%%%%%%%%%%%%%%%%%%%%%%%%%%%%%%%%%%%%%%%%%%%%%
\clearpage
\section*{Figure Captions}
%%%%%%%%%%%%%%%%%%%%%%%%%%%%%%%%%%%%%%%%%%%%%%%%%%%%%
%==============================================================
%==============================================================
%==============================================================
%==============================================================

%==============================================================
%==============================================================
%==============================================================
%==============================================================

%==============================================================
%==============================================================

%==============================================================
%==============================================================

%%%%%%%%%%%%%%%%%%%%%%%%%%%%%%%%%%%%%%%%%%%%%%%%%%%%%

\begin{thebibliography}{99}
\bibitem{footnote1}
Let us denote by $u_i$ and $v_i$ the initial and final
velocities of the $i$th object, respectively. Then conservation
of energy and momentum require $u_1+u_2=v_1+v_2$ and 
$u_1^2+u_2^2=v_1^2+v_2^2$. These coupled equations have a unique set 
of solutions: $v_1=u_2$ and $v_2=u_1$. If they do not make a collision,
then $v_1=u_1$ and $v_2=u_2$.
\bibitem{footnote1x}
Let us consider the last collision between $C$ and $B$ 
at $Q_k$ when
$\alpha=\alpha_k^{\textrm{magic}}$.
At $Q_k$ the velocity of $C$ changes from $v$ to $0$
and that of $B$ changes from $V'$ to $V$.
Energy-momentum conservation
requires $\alpha v+V'=V$ and $\alpha v^2+V'^2=V^2$.
The solutions to $v$ and $V'$ are
$v=2V/(1+\alpha)$ and $V'=V(1-\alpha)/(1+\alpha)$.
In a similar manner, we find that
the velocities of $A$ and $C$ must be
$4\alpha V/(1+\alpha)^{2}$ and $-2V(1-\alpha)/(1+\alpha)^{2}$, 
respectively, just before the collision at $P_{k}$.
This guarantees that $A$ stops at $P_{k}$ and $C$ moves with
the velocity $v=2V/(1+\alpha)$ between $P_{k}$ and $Q_{k}$.
\bibitem{footnote3}
We can show that
$-\alpha V/(2+\alpha)\le a_t\le 0$ by making use of  
Eq.~(\ref{eq:a-terminal})
and the constraint 
$\pi-(\psi/2)\le N_A<\pi+(\psi/2)$ 
that derives from Eq.~(\ref{eq:NA-ans}).
\bibitem{footnote2}
Note that $\alpha=0$ represents the case that the ball is absent.
\end{thebibliography}
\end{document}